\documentclass[11pt]{article}

\usepackage{amsmath}
\usepackage{amsfonts}
\usepackage{amssymb}
\usepackage{graphicx}
\usepackage[table,xcdraw]{xcolor}
\usepackage{orcidlink}
\usepackage{hyperref}
\hypersetup{colorlinks=true, linkcolor=red, citecolor=red, urlcolor=magenta}
\usepackage{xstring}
\newcommand{\hrefbibentry}[2]{%
  \IfStrEq{#1}{}{#2}{%
    \IfBeginWith{#1}{http}%
      {\href{#1}{\textcolor{blue}{#2}}}%
      {\href{https://doi.org/#1}{\textcolor{blue}{#2}}}}}
\usepackage{float}
\usepackage{multirow}
\usepackage{array}
\usepackage{longtable}
\usepackage{booktabs}
\usepackage{colortbl}
\usepackage{tabularx}
\usepackage{xltabular}
\usepackage{caption}

\newcolumntype{C}{>{\centering\arraybackslash}X}
\newcolumntype{L}{>{\raggedright\arraybackslash}X}
\newcolumntype{R}{>{\raggedleft\arraybackslash}X}
\usepackage{geometry}
\renewcommand{\arraystretch}{1.6}
\usepackage[numbers,sort&compress]{natbib}
\numberwithin{equation}{section}
\newcommand{\dd}{\mathrm{d}}
\newcommand{\rH}{r_{h}}
\newcommand{\rps}{r_{\rm ph}}

\newcommand{\Meff}{M_{\rm eff}}
\newcommand{\Qeff}{Q_{\rm eff}}
\newcommand{\Msun}{M_{\odot}}

\begin{document}
\begin{center}

{\LARGE\bfseries Constraints on Scalar--Tensor--Vector Gravity Theory Parameters Inferred from Quasiperiodic Oscillations\par}

\vspace{16pt}

{\large Marco Muccino$^{1,2,3,4}$ \orcidlink{0000-0002-2234-9225}, Kuantay Boshkayev$^{2,5}$ \orcidlink{0000-0002-1385-270X}, Binay Prakash Akhouri$^{6}$\,\orcidlink{0000-0003-4978-9367},\\[2pt] Izzet Sakall{\i}$^{7}$ \orcidlink{0000-0001-7827-9476} and Yassine Sekhmani$^{2,8,9}$\,\orcidlink{0000-0001-7448-4579}\par}

\vspace{10pt}

\begin{minipage}{0.90\textwidth}
\small\itshape\centering
$^{1}$Universit\`a di Camerino, Divisione di Fisica, Via Madonna delle carceri 9, 62032 Camerino, Italy\\
$^{2}$Al-Farabi Kazakh National University, Al-Farabi Ave.\ 71, 050040 Almaty, Kazakhstan\\
$^{3}$INAF, Catania Astrophysical Observatory, Via S.\ Sofia 79, 95123 Catania, Italy\\
$^{4}$ICRANet, Piazza della Repubblica 10, 65122 Pescara, Italy\\
$^{5}$Kazakh-British Technical University, Tole Bi Str.\ 59, 050000 Almaty, Kazakhstan\\
$^{6}$Department of Physics, Suraj Singh Memorial College,Ranchi University, Ranchi 834008, Jharkhand, India\\
$^{7}$Physics Department, Eastern Mediterranean University, Famagusta 99628, Cyprus\\
$^{8}$Center for Theoretical Physics, Khazar University, 41 Mehseti Street, Baku AZ1096, Azerbaijan\\
$^{9}$Fesenkov Astrophysical Institute, Observatory 23, 050020 Almaty, Kazakhstan
\end{minipage}

\vspace{8pt}

{\footnotesize\ttfamily marco.muccino@unicam.it \quad kuantay@mail.ru \quad binayakhouri@yahoo.in\\
izzet.sakalli@emu.edu.tr \quad sekhmaniyassine@gmail.com}

\end{center}

\vspace{10pt}

\begin{center}
\begin{minipage}{0.93\textwidth}
{\bfseries Abstract.}\ \ We study circular geodesics of neutral test particles in the static charged
solution of Scalar--Tensor--Vector Gravity (STVG), and we use the twin kilohertz quasiperiodic oscillations (QPOs) of twelve accreting compact objects to bound its parameters. Starting from the effective potential we obtain closed forms for the specific energy, the specific angular momentum and the Keplerian angular velocity, and from these the radial and vertical epicyclic frequencies.
The horizon, photon sphere, shadow radius and innermost stable circular orbit are given in closed form as well, and each one reduces to the RN and Schwarzschild value in the
appropriate limit. Null geodesics are integrated numerically and show how the enhanced coupling widens the capture cross section while leaving the logarithmic divergence of the deflection angle at
the photon sphere intact. Within the RP model we then run Metropolis--Hastings MCMC simulations on the QPO pairs of eight neutron stars and four microquasars, comparing the Schwarzschild spacetime, the RN spacetime, STVG with vanishing charge and the full STVG solution. The best fits are obtained mainly for the charged solutions in the neutron star sample, while all four models describe the black hole sample equally well. We show
that the orbital dynamics depends on the mass $M$, the coupling $\alpha$ and the charge $Q$ only through the two combinations $\Meff=(1+\alpha)M$ and $\Qeff^{2}=(1+\alpha)(\alpha M^{2}+Q^{2})$, which accounts for the multimodal posteriors found in the neutron star sample and sets a model-independent limit on what QPO timing alone can measure. The astrophysical implications of
these bounds, in particular for inferred masses above the Tolman--Oppenheimer--Volkoff limit, are discussed in detail.

\vspace{6pt}
\end{minipage}
\end{center}

\vspace{12pt}

\section{Introduction}\label{isec1}

General relativity (GR) remains the reference description of gravitation on scales from the
laboratory to the cosmological horizon, and the direct detection of gravitational waves together
with the resolved images of supermassive compact objects has extended its validated domain into the
strong field \citep{Abbott:2016,Abbott:2017,Abbott:2016a,Abbott:2017a,Akiyama:2019}. That success is
not the same thing as a proof of uniqueness. Binary mergers, the innermost regions of accretion
flows and the environments of galactic nuclei probe curvature regimes in which classical or quantum
corrections to the Einstein--Hilbert action would first become visible, and the observational
program that follows from this is to build alternative theories whose predictions differ from GR
by an amount that current instruments can resolve \citep{Clifton:2012,Johannsen:2016}. A useful
alternative must therefore do two things at once. It must reproduce the weak-field tests that GR
already passes, and it must carry a small number of free parameters that leave a measurable
imprint somewhere in the strong field.

Scalar--Tensor--Vector Gravity (STVG), usually called Modified Gravity or MOG, is one such theory.
Moffat introduced it as a covariant extension of GR in which a massive vector field is added to the
metric sector, with scalar fields promoting the gravitational coupling and the vector mass to
dynamical quantities \citep{Moffat:1995,Moffat:2005,Moffat:2006}. The enhancement of the
gravitational constant is controlled by a dimensionless parameter $\alpha$ through
$G=G_{N}(1+\alpha)$, while the repulsive Yukawa force sourced by the vector field opposes that
enhancement at short range. The theory was constructed to account for galactic rotation curves,
gravitational lensing by clusters and cluster dynamics without a particulate dark sector, and it has
been tested against those data sets with some success \citep{Mourelle:2024,Frenk:1996,Harikumar:2022,Bartelmann:2010,Carroll:2001,Gupta:2025}.
Its black hole solutions have been studied in their own right: the static and rotating cases, their shadows, their innermost stable circular orbits and the motion of neutral and charged particles around them \citep{Moffat:2015,Lee:2017,Sharif:2018,Rayimbaev:2021,Nishonov:2025,Oteev:2026}. What
matters for the present work is that the charged static solution carries three parameters, $M$,
$\alpha$ and $Q$, and that these enter the orbital frequencies of the accretion flow in a way that
timing observations can in principle disentangle.

The observational handle we use is the quasiperiodic oscillation (QPO). QPOs appear as narrow peaks
in the power density spectra of the X-ray light curves of accreting neutron stars and black holes,
and were identified as a diagnostic of the innermost accretion flow soon after their discovery
\citep{Hameury:1985,Lewin:1988,Klis:1989}. The signal is understood to originate in gas
spiraling inward through the disc: as material accumulates in a localized region close to the
compact object it is heated strongly enough to radiate in X-rays, and the modulation repeats on a
timescale set by the orbital motion. Because the repetition is not strictly periodic the features
are called quasiperiodic. Their frequencies run from a few hertz to above one kilohertz, and the
high-frequency branch above $0.1$ kHz falls squarely in the range of the Keplerian and epicyclic
frequencies at a few gravitational radii. Twin-peak kilohertz QPOs, in which two narrow peaks appear
simultaneously and drift together, are the most constraining subclass, and they have been detected
in a large number of low-mass X-ray binaries
\citep{Ford:1997,Wijnands:1998,Mendez:1998,Mendez:1999,Jonker:2000,Mendez:2000,Homan:2002,Jonker:2002,Boutloukos:2006,Boutelier:2009,Buisson:2019,Cherepashchuk:2021}.

Several mechanisms have been proposed to convert orbital motion into an observed frequency pair. The
relativistic precession (RP) model identifies the upper peak with the Keplerian frequency and the
lower peak with the periastron precession frequency, that is, with the difference between the
Keplerian and radial epicyclic frequencies \citep{Stella:1998,Stella:1999,Stella:1999a}.
Diskoseismic models attribute the peaks to trapped oscillation modes of the disc itself
\citep{Wagoner:1999,Wagoner:2001}, and resonance models invoke a nonlinear coupling between the
radial and vertical epicyclic oscillations at a rational frequency ratio, most often $3\!:\!2$
\citep{Abramowicz:2001,Kluzniak:2001,Abramowicz:2003,Abramowicz:2004,Rebusco:2004,Torok:2005}. The
three families make different predictions for the shape of the $f_{\rm L}(f_{\rm U})$ relation, and
comparisons among them have been carried out for both neutron star and black hole sources
\citep{Urbanec:2010,Torok:2010,Torok:2012,Torok:2016,Smith:2021,Motta:2024}. We adopt the RP model
here because it involves no free parameter beyond the spacetime itself, so that any preference the
data express is a statement about the metric rather than about a coupling coefficient inserted by
hand.

Once a mechanism is fixed, the frequency pair becomes a probe of the geometry. This program has
been carried out for a wide range of spacetimes: for rotating and quadrupole-deformed neutron star
exteriors in the Hartle--Thorne approximation \citep{Hartle:1967,Hartle:1968,Baubock:2013,Urbanec:2013,Urbancova:2019,Boshkayev:2014,Boshkayev:2015,Boshkayev:2016,Boshkayev:2024,Boshkayev:2024a,Boshkayev:2026},
for regular black holes, for hairy and ultracompact objects, and for a variety of modified-gravity
backgrounds
\citep{Bakala:2010,GondekRosinska:2014,Doneva:2014,Staykov:2015,Vieira:2017,Danchev:2020,Shaymatov:2020,Rayimbaev:2021a,Delgado:2022,Falco:2023,Shaymatov:2023,Mustafa:2025,Gurmani:2026}.
Constraints on specific theories have been extracted in the same way
\citep{Bambi:2012,Bambi:2016,Rayimbaev:2022,Rahmatov:2024,Khan:2024,Xamidov:2025,Shermatov:2025},
and recent work has combined QPO timing with shadow and thermodynamic information for quantum
corrected and dark-matter-dressed black holes
\citep{AlBadawi:2026,Ahmed:2026,Ahmed:2026a,Ahmed:2026b,Sucu:2025reg,Sucu:2026pla,Sekhmani:2025mod,Ahmed:2025her,Sakalli:2026zit}.
Most of these studies treat a single spacetime parameter, or treat two parameters that enter the
lapse function through independent powers of $1/r$. The charged STVG solution is different in a way
that turns out to matter: its two corrections enter at the same order, and this has consequences
for what the data can and cannot determine.

Eight neutron star low-mass X-ray binaries and four microquasars are analyzed here. The neutron star
sample, Cir X-1, GX 5--1, GX 17+2, GX 340+0, Sco X-1, 4U 1608--52, 4U 1728--34 and 4U 0614+091,
splits according to accretion rate and to the track traced on the color--color diagram. The Z
sources GX 5--1, GX 17+2, GX 340+0 and Sco X-1, together with the strongly variable Cir X-1, accrete
at high rates and show horizontal branch oscillations as their low-frequency QPOs; the atoll sources
4U 1608--52, 4U 1728--34 and 4U 0614+091 accrete more slowly and display clean twin kilohertz peaks
\citep{Torok:2016,Boshkayev:2026}. The black hole sample consists of XTE J1550--564, GRO J1655--40,
GRS 1915+105 and H1743--322. A recent analysis of the same eight neutron stars in the Hartle--Thorne
approximation established that an exterior metric carrying mass, slow rotation and a mass quadrupole
represents these objects accurately, and that the rotational deformation is small enough for the
approximation to remain reliable when physical parameters are extracted \citep{Boshkayev:2026}. That
result is the baseline against which a static modified-gravity description should be judged.

Our aim is threefold. First, we give a complete analytic treatment of the charged STVG geometry at
the level required by the RP model, including the horizon structure, the curvature invariants, the
effective source and its energy conditions, the photon sphere and shadow, the null geodesic
structure, the circular orbit parameters, the ISCO and both epicyclic frequencies, checking every
expression against its Reissner--Nordstr\"om (RN) and Schwarzschild limits. Second, we fix the radial epicyclic frequency by three independent
checks, since the Schwarzschild limit alone does not determine its $r^{-5}$ and $r^{-6}$
coefficients. Third, we perform the Markov chain Monte Carlo (MCMC) analysis and interpret the resulting bounds, paying particular attention to an exact parameter degeneracy that the geometry itself imposes and that explains the multimodal posteriors found in the neutron star
sample.

The spacetime signature is $(-,+,+,+)$ and geometric units with $G_{N}=c=1$ are used throughout the
analytic sections; quantities with direct astrophysical meaning are quoted in physical units, with
masses in solar masses and lengths in kilometers, using $G_{N}\Msun/c^{2}=1.47663$ km. The paper is
organized as follows. Section \ref{isec2} presents the charged STVG solution, its horizon structure,
its curvature invariants and the effective source that supports it, and establishes the exact map
onto RN that governs the rest of the analysis. Section \ref{isec3} treats null geodesics, the photon
sphere, the shadow and the deflection angle. Section \ref{isec4} derives the circular orbit
parameters and the ISCO. Section \ref{isec5} obtains the epicyclic frequencies, corrects the radial
one and assembles the RP model prediction. Section \ref{isec6} describes the likelihood, the priors
and the model selection criterion. Section \ref{isec7} presents the MCMC results for the twelve
sources and discusses them. Section \ref{isec8} closes with our conclusions.

\section{The charged black hole in scalar tensor vector gravity}\label{isec2}

Before turning to orbits we fix the geometry and establish what kind of matter distribution supports
it, because both questions bear directly on the interpretation of the fits. The static charged STVG
solution looks like a RN metric with shifted coefficients, and we show below that
this resemblance is exact rather than superficial: the two geometries are related by a redefinition
of mass and charge. That statement fixes in advance which combinations of $(M,\alpha,Q)$ any
orbital observable can depend on, and it will reappear in Section \ref{isec7} as the explanation of
the multiple parameter modes recovered by the sampler. We also record the curvature invariants and
the effective stress tensor, which together delimit the region of parameter space in which the
solution is physically acceptable.

The STVG action contains the Einstein--Hilbert term, a massive vector field $\phi^{\mu}$ with field
strength $B_{\mu\nu}=\partial_{\mu}\phi_{\nu}-\partial_{\nu}\phi_{\mu}$, and scalar fields
governing the running of the gravitational coupling $G$ and of the vector mass. In the regime
relevant here the scalars are treated as slowly varying, the gravitational coupling takes the
enhanced value $G=G_{N}(1+\alpha)$, and the vector field carries a gravitational source charge
proportional to the mass, $Q_{g}=\sqrt{\alpha G_{N}}\,M$, together with an ordinary Maxwellian charge
$Q$ \citep{Moffat:2006,Moffat:2015}. Because the electromagnetic sector and the vector sector both
contribute at order $r^{-2}$, the charged solution has features that neither of its limits displays
on its own \citep{Dain:2010}. The resulting static spherically symmetric line element is
\citep{Sucu:2025}
\begin{equation}
\dd s^{2}=-f(r)\,\dd t^{2}+\frac{\dd r^{2}}{f(r)}+r^{2}\!\left(\dd\theta^{2}+\sin^{2}\theta\,\dd\varphi^{2}\right),
\label{eq:metric}
\end{equation}
with the lapse
\begin{equation}
f(r)=1-\frac{2(1+\alpha)M}{r}+\frac{(1+\alpha)\left(\alpha M^{2}+Q^{2}\right)}{r^{2}}.
\label{eq:lapse}
\end{equation}
Here $M$ is the total mass of the compact object, $\alpha$ is the dimensionless deviation parameter,
and $Q$ is the electric or magnetic charge. Setting $\alpha=0$ returns the RN solution and setting
$\alpha=Q=0$ returns Schwarzschild; setting $Q=0$ alone leaves the Schwarzschild--MOG solution, whose
$r^{-2}$ coefficient $\alpha(1+\alpha)M^{2}$ is exactly the combination
$G G_{N}\alpha M^{2}$ produced by the vector source charge. The two corrections therefore act in
opposition: the numerator of the $r^{-1}$ term is enhanced by $(1+\alpha)$, deepening the potential
well, while the $r^{-2}$ term is repulsive and dominates at small radius.

It is convenient throughout to introduce
\begin{equation}
A\equiv 1+\alpha,\qquad C\equiv \alpha M^{2}+Q^{2},\qquad x\equiv \frac{r}{M},\qquad q\equiv\frac{Q}{M},\qquad c\equiv \alpha+q^{2},
\label{eq:shorthand}
\end{equation}
so that $f=1-2A/x+Ac/x^{2}$ in units $M=1$. Comparing Eq.~(\ref{eq:lapse}) with the RN lapse
$1-2\Meff/r+\Qeff^{2}/r^{2}$ gives an exact correspondence,
\begin{equation}
\Meff=(1+\alpha)M,\qquad
\Qeff^{2}=(1+\alpha)\left(\alpha M^{2}+Q^{2}\right)=A\,C .
\label{eq:map}
\end{equation}
The charged STVG geometry is therefore isometric to a RN geometry of mass $\Meff$
and charge $\Qeff$. Nothing computed from the metric alone can distinguish the two, and every
geodesic observable is a function of $\Meff$ and $\Qeff$ only. Since the three physical parameters
$(M,\alpha,Q)$ enter through two combinations, one direction in parameter space is exactly flat for
any measurement built on test-particle motion. We return to the observational consequences in
Section \ref{isec7}; for the moment we note only that this is a property of the solution and not an
artifact of the fitting procedure.

The horizons follow from $f(\rH)=0$. Solving the quadratic and simplifying the discriminant,
\begin{equation}
r_{\pm}=(1+\alpha)M\pm\sqrt{(1+\alpha)\left(M^{2}-Q^{2}\right)},
\label{eq:horizons}
\end{equation}
a compact result whose structure deserves comment. The discriminant factorizes as
$A^{2}M^{2}-AC=A(A M^{2}-\alpha M^{2}-Q^{2})=A(M^{2}-Q^{2})$, so the $\alpha$ dependence cancels
inside the square root except for the overall factor $A$. An event horizon exists provided
$\alpha>-1$ and $|Q|\le M$, and the extremal configuration occurs at $|Q|=M$ regardless of the value
of $\alpha$. This is a nontrivial statement: the STVG coupling rescales both horizons but cannot by
itself expose a naked singularity, and it cannot prevent one either. For $Q=0$ we recover
$r_{\pm}=(1+\alpha)M\pm M\sqrt{1+\alpha}$, the Schwarzschild--MOG horizons, and for $\alpha=0$ the
RN pair $M\pm\sqrt{M^{2}-Q^{2}}$. Setting both to zero gives $r_{+}=2M$.

\begin{figure}[t!]
\centering
\includegraphics[width=\textwidth]{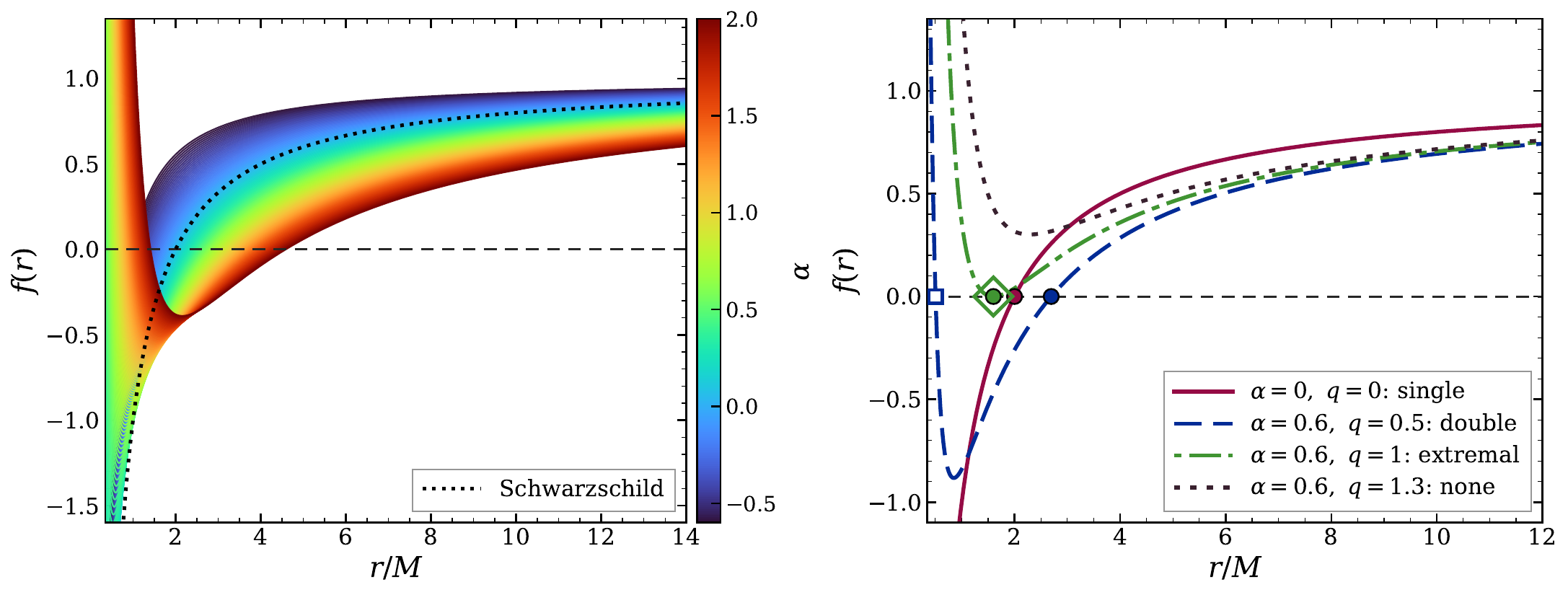}
\vspace{-0.3cm}\\
\caption{Left: lapse function of Eq.~(\ref{eq:lapse}) drawn as a dense continuous family at fixed
$q=Q/M=0.4$, with the STVG coupling swept over $-0.6\le\alpha\le 2$ and encoded by the ribbon color;
the dotted black curve is the Schwarzschild lapse and the dashed horizontal line marks $f=0$. Right:
the four horizon configurations permitted by Eq.~(\ref{eq:horizons}). Filled circles mark the outer
root $r_{+}$, open squares the inner root $r_{-}$, and the open diamond the degenerate root of the
extremal case. The solid curve has a single horizon because $r_{-}=0$, the dashed curve has two
distinct horizons, the dot-dashed curve is extremal with $r_{+}=r_{-}=(1+\alpha)M$, and the dotted
curve stays positive everywhere so that no horizon forms.\\}
\label{fig:lapse}
\end{figure}

The left panel of Fig.~\ref{fig:lapse} displays the one-parameter family generated by the coupling.
The rightward migration of the zero crossing with increasing $\alpha$, set against its leftward
migration with increasing charge, are the two competing tendencies that the QPO fits will later have
to arbitrate.

The right panel sorts the solutions by horizon multiplicity, and the classification is worth stating
explicitly because it controls which regions of parameter space describe a black hole at all.
Equation~(\ref{eq:horizons}) is a quadratic in $r$ whose discriminant is $(1+\alpha)(M^{2}-Q^{2})$,
so the count of roots is governed by $q$ alone once $\alpha>-1$ is imposed. When $q=0$ and $\alpha=0$
the inner root sits at the origin and the geometry carries a single horizon at $r_{+}=2M$, the
Schwarzschild case. Switching on either parameter separates the roots: for $0<q<1$, and equally for
$\alpha>0$ at $q=0$, the lapse dips below zero over a finite interval and two distinct horizons
appear, an outer event horizon and an inner Cauchy horizon at
$r_{-}=(1+\alpha)M-\sqrt{(1+\alpha)(M^{2}-Q^{2})}$. Note that a nonzero coupling alone is enough to
produce an inner horizon, since $r_{-}=(1+\alpha)M-M\sqrt{1+\alpha}>0$ for $\alpha>0$; the vector
source charge $Q_{g}=\sqrt{\alpha G_{N}}M$ plays the same structural role in the Schwarzschild--MOG
solution that the Maxwellian charge plays in RN. At $q=1$ the discriminant
vanishes and the two roots merge into the degenerate value $r_{+}=r_{-}=(1+\alpha)M$, an extremal
configuration at which the lapse touches zero tangentially and the surface gravity vanishes. For
$q>1$ the lapse stays positive for all $r>0$ and no horizon forms, leaving a naked singularity.
The threshold is at $|Q|=M$ for every value of $\alpha$, so the coupling rescales all three radii
without shifting the boundary between the black hole and horizonless branches. This last point
matters for Section \ref{isec7}: several of the neutron star fits return $|Q|>M$ in the recovered
parameters, which is admissible for a stellar surface but would be unacceptable for a black hole. The curvature of the geometry
is characterized by the Ricci scalar and the Kretschmann invariant. Direct computation from
Eq.~(\ref{eq:metric}) gives
\begin{equation}
R=0,
\label{eq:ricci}
\end{equation}
and
\begin{equation}
K=R_{\mu\nu\rho\sigma}R^{\mu\nu\rho\sigma}
=\frac{8(1+\alpha)^{2}}{r^{8}}\left[\,7\alpha^{2}M^{4}-12\alpha M^{3}r+14\alpha M^{2}Q^{2}
+6M^{2}r^{2}-12MQ^{2}r+7Q^{4}\,\right].
\label{eq:kret}
\end{equation}
Setting $\alpha=0$ reduces Eq.~(\ref{eq:kret}) to
$48M^{2}/r^{6}-96MQ^{2}/r^{7}+56Q^{4}/r^{8}$, the standard RN invariant, and setting $Q=0$ as well
gives $48M^{2}/r^{6}$. The invariant diverges only at $r=0$, so the central singularity is the sole
curvature singularity and the horizons of Eq.~(\ref{eq:horizons}) are regular. In terms of the
effective parameters, Eq.~(\ref{eq:kret}) is nothing but the RN invariant with $M\to\Meff$ and
$Q^{2}\to\Qeff^{2}$, as the isometry of Eq.~(\ref{eq:map}) requires.

The vanishing Ricci scalar already indicates that whatever supports the geometry is traceless.
Computing the Einstein tensor and reading off the effective stress tensor
$T^{\mu}{}_{\nu}=G^{\mu}{}_{\nu}/8\pi$ in the static orthonormal frame yields
\begin{equation}
\rho=-T^{t}{}_{t}=\frac{(1+\alpha)\left(\alpha M^{2}+Q^{2}\right)}{8\pi r^{4}},\qquad
p_{r}=-\rho,\qquad p_{\theta}=p_{\varphi}=+\rho ,
\label{eq:source}
\end{equation}
which is the anisotropic, traceless, Maxwell-like profile familiar from RN with the replacement
$Q^{2}\to\Qeff^{2}$. Two conditions follow. The weak and null energy conditions require
$\rho\ge 0$, hence
\begin{equation}
\alpha>-1\qquad\text{and}\qquad \alpha M^{2}+Q^{2}\ge 0 ,
\label{eq:ec}
\end{equation}
the second of which permits moderately negative couplings provided the Maxwellian charge is large
enough to compensate, $Q^{2}\ge|\alpha|M^{2}$. The strong energy condition is satisfied wherever the
weak one is, since $\rho+\sum_{i}p_{i}=2\rho\ge 0$. Configurations with $\alpha<-1$ invert the sign
of the gravitational coupling and are discarded. Both bounds in Eq.~(\ref{eq:ec}) are comfortably
respected by the parameter ranges recovered in Section \ref{isec7}, the most negative coupling in the
neutron star sample being $\alpha\simeq-0.48$ for GX 340+0.
\section{Null geodesics, photon sphere and the shadow}\label{isec3}

Massless particles do not enter the RP model directly, but they fix the two scales against which the
timing results must be read: the photon sphere, which is the innermost circular null orbit and hence
the boundary of the region from which any radiation can escape, and the critical impact parameter,
which sets the apparent size of the object on the sky. Both are sensitive to $\alpha$ and $Q$ in the
same way as the orbital frequencies, so a spacetime that improves the QPO fit necessarily predicts a
shifted shadow, and the two constraints can be confronted with each other. In this section we obtain
the null structure analytically and then integrate the trajectories to display the lensing pattern.

For a photon confined to the equatorial plane the conserved energy $E=f\dot t$ and angular momentum
$L=r^{2}\dot\varphi$ combine with the null condition to give
\begin{equation}
\left(\frac{\dd r}{\dd\lambda}\right)^{2}=E^{2}-\frac{L^{2}}{r^{2}}f(r),
\qquad
\left(\frac{\dd u}{\dd\varphi}\right)^{2}=\frac{1}{b^{2}}-u^{2}f\!\left(\frac{1}{u}\right)\equiv F(u),
\label{eq:null}
\end{equation}
where $u=1/r$ and $b=L/E$ is the impact parameter. Circular null orbits satisfy
$rf'(r)=2f(r)$, equivalently $\left(f/r^{2}\right)'=0$, which for Eq.~(\ref{eq:lapse}) reduces to the
quadratic
\begin{equation}
r^{2}-3(1+\alpha)Mr+2(1+\alpha)\left(\alpha M^{2}+Q^{2}\right)=0 ,
\label{eq:phquad}
\end{equation}
whose outer root is the photon sphere,
\begin{equation}
\rps=\frac{3(1+\alpha)M+\sqrt{9(1+\alpha)^{2}M^{2}-8(1+\alpha)\left(\alpha M^{2}+Q^{2}\right)}}{2}.
\label{eq:rph}
\end{equation}
For $\alpha=Q=0$ this gives $\rps=3M$ and for $\alpha=0$ it gives
$\left[3M+\sqrt{9M^{2}-8Q^{2}}\right]/2$, the RN photon sphere. The polynomial in
Eq.~(\ref{eq:phquad}) is worth pausing on, because the same expression will appear in
Section \ref{isec4} inside the square root that normalizes the specific energy and angular momentum
of circular timelike orbits. Its vanishing is precisely the statement that those quantities diverge,
as they must at the photon sphere, and this coincidence provides an internal consistency check on
the timelike sector that we exploit below.

The critical impact parameter, which for a distant observer equals the shadow radius, is
\begin{equation}
b_{\rm c}=R_{\rm sh}=\frac{\rps}{\sqrt{f(\rps)}} .
\label{eq:bcrit}
\end{equation}
Evaluating Eq.~(\ref{eq:bcrit}) for Schwarzschild returns $3\sqrt{3}\,M\simeq5.196M$, and our
numerical implementation reproduces this to machine precision. The instability timescale of the
circular null orbit is set by the principal Lyapunov exponent
\begin{equation}
\lambda_{\rm L}=\sqrt{\frac{f(\rps)}{2}\left[\frac{2f(\rps)}{\rps^{2}}-f''(\rps)\right]},
\label{eq:lyap}
\end{equation}
which controls the spacing of successive photon-ring images and, through the eikonal correspondence,
the damping rate of high-multipole quasinormal modes. For Schwarzschild
Eq.~(\ref{eq:lyap}) gives $M\lambda_{\rm L}=1/(3\sqrt{3})\simeq0.1925$, again reproduced by our
implementation.

\begin{figure}[ht!]
\centering
\includegraphics[width=\textwidth]{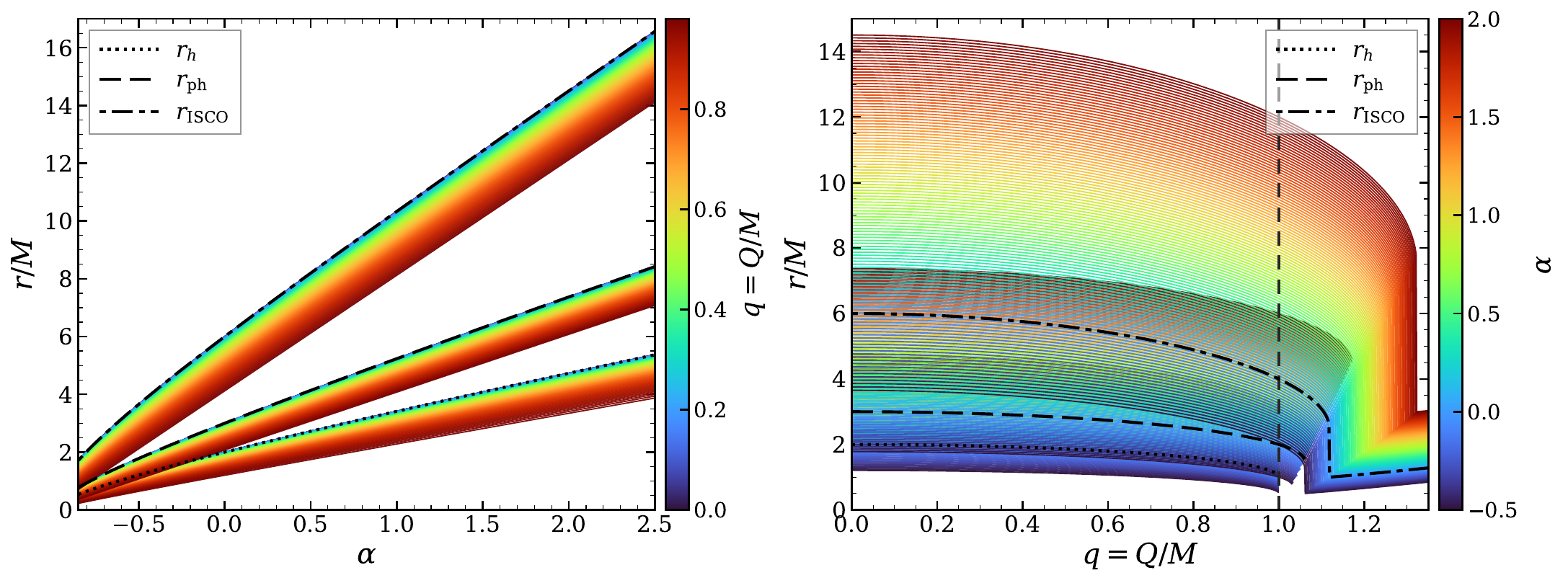}
\vspace{-0.3cm}\\
\caption{Characteristic radii of the charged STVG spacetime. Three ribbon families are drawn in each
panel and are, from bottom to top at fixed abscissa, the event horizon $r_{h}$, the photon sphere
$\rps$ of Eq.~(\ref{eq:rph}) and the innermost stable circular orbit $r_{\rm ISCO}$ obtained from
Eq.~(\ref{eq:iscocubic}). In the left panel the abscissa is the coupling and the ribbon color
encodes $q=Q/M$ over $0\le q\le0.98$; in the right panel the abscissa is the reduced charge and the
color encodes $\alpha$ over $-0.5\le\alpha\le2$. Black curves in each panel repeat the three radii
for the uncharged, respectively uncoupled, case with the line styles given in the legend. The
vertical dashed line in the right panel marks the extremal value $q=1$, beyond which
Eq.~(\ref{eq:horizons}) has no real root. The vertical range is truncated at $r=17M$ on the left
and at $r=15M$ on the right so that the three families remain resolved.}
\label{fig:radii}
\end{figure}

Figure \ref{fig:radii} collects the three characteristic radii. The ordering
$r_{+}<\rps<r_{\rm ISCO}$ is preserved throughout the physically admissible region, and all three
grow roughly linearly with $\alpha$ at fixed charge. The growth is close to proportional to
$(1+\alpha)$, as Eq.~(\ref{eq:map}) demands: at $q=0$ the three radii are exactly $(1+\alpha)$ times
their Schwarzschild values only when $\alpha=0$, because $\Qeff$ also grows with $\alpha$, and the
mild departure from proportionality visible in the figure is the charge contribution.

The null trajectories themselves follow from integrating Eq.~(\ref{eq:null}). Writing $u_{t}$ for the
periastron root of $F(u)=0$, the deflection accumulated between infinity and periastron is
\begin{equation}
\Delta\varphi=\int_{0}^{u_{t}}\frac{\dd u}{\sqrt{F(u)}},\qquad
\hat\alpha_{\rm defl}=2\Delta\varphi-\pi .
\label{eq:defl}
\end{equation}
The integrand has an inverse square-root singularity at $u_{t}$, which we remove with the
substitution $u=u_{t}-t^{2}$; the transformed integrand tends to $2/\sqrt{|F'(u_{t})|}$ as $t\to0$
and is bounded throughout. Applied to Schwarzschild the resulting quadrature returns
$\hat\alpha_{\rm defl}=0.03962$ rad at $b=20\,b_{\rm c}$, against the weak-field expansion
$4M/b+15\pi M^{2}/(4b^{2})=0.03958$ rad, and $6.5106$ rad at $b=1.001\,b_{\rm c}$, against the
strong-deflection value $-\ln(b/b_{\rm c}-1)+\ln\!\left[216(7-4\sqrt3)\right]-\pi=6.5069$ rad. Both
limits are recovered, so the integrator is trustworthy across the full range of impact parameters.

\begin{figure}[ht!]
\centering
\includegraphics[width=0.92\textwidth]{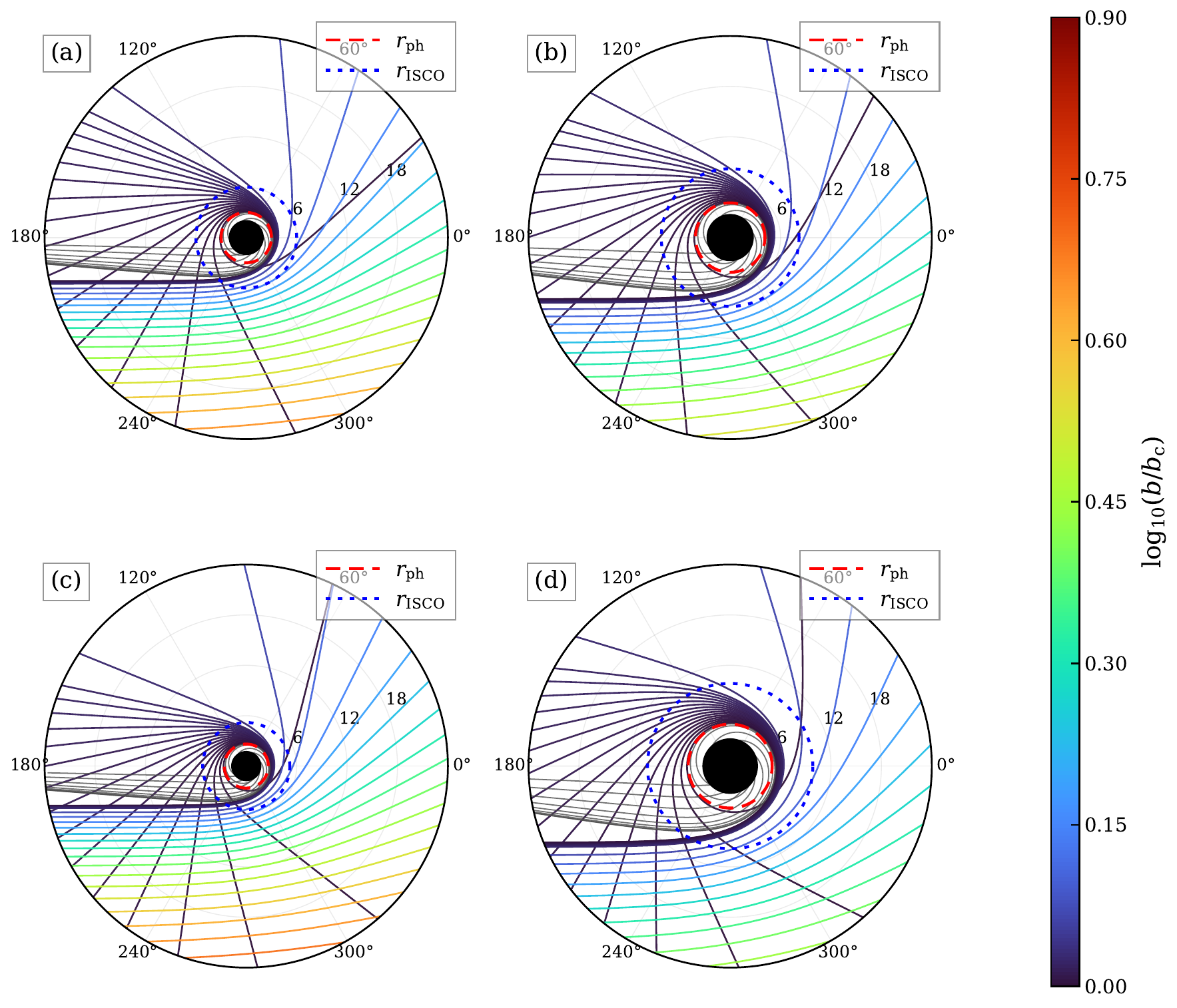}
\caption{Equatorial null geodesics of the charged STVG spacetime, obtained by integrating
Eq.~(\ref{eq:null}) with the periastron substitution described in the text. Rays enter from the left
along $\varphi=\pi$ with impact parameters ranging from $b=1.0006\,b_{\rm c}$ to $b=8\,b_{\rm c}$,
the ribbon color encoding $\log_{10}(b/b_{\rm c})$ over exactly the range of impact
parameters drawn; grey curves are captured rays with
$b<b_{\rm c}$ that terminate on the horizon, shown as the filled black disc. The dashed red circle is
the photon sphere of Eq.~(\ref{eq:rph}) and the dotted blue circle is the ISCO. The four cases are
$(\alpha,q)=(0,0)$ in panel (a), $(0.5,0)$ in panel (b), $(0,0.7)$ in panel (c) and $(1,0.5)$ in
panel (d), and the radial scale is common to all four, so that the growth of the capture region with
$\alpha$ and its contraction with $q$ can be read directly. Near-critical rays wind repeatedly about
the photon sphere before escaping, which is the geometrical origin of the logarithmic divergence in
Eq.~(\ref{eq:defl}).}
\label{fig:geodesics}
\end{figure}

Figure \ref{fig:geodesics} shows the lensing pattern for four representative parameter choices. The
qualitative structure is the same in all four panels: rays with $b$ well above $b_{\rm c}$ are bent
by a small angle and leave along nearly straight asymptotes, rays approaching $b_{\rm c}$ from above
wind several times around the photon sphere before escaping into a direction essentially uncorrelated
with their entry direction, and rays below $b_{\rm c}$ spiral inward and cross the horizon. What
changes between panels is the scale. Comparing panel (a) with panel (b), raising $\alpha$ from zero
to one half expands the horizon, the photon sphere and the capture cross section by roughly a third,
so that a ray which escapes in the Schwarzschild geometry can be captured in the STVG one at the same
$b/M$. Comparing panel (a) with panel (c), adding charge at fixed coupling has the opposite effect
and shrinks all three scales, consistent with the repulsive character of the $r^{-2}$ term. Panel (d)
combines a large coupling with a moderate charge and shows that the coupling wins: the capture
region is the largest of the four despite $q=0.5$.

\begin{table}[ht!]
\centering
\footnotesize
\setlength{\tabcolsep}{15pt}
\renewcommand{\arraystretch}{1.55}
\begin{tabular}{|r|r|r|r|r|r|r|}
\hline
$\boldsymbol{\alpha}$ & $\boldsymbol{q}$ & $\boldsymbol{r_{h}/M}$ & $\boldsymbol{\rps/M}$ & $\boldsymbol{R_{\rm sh}/M}$ & $\boldsymbol{r_{\rm ISCO}/M}$ & $\boldsymbol{M\lambda_{\rm L}}$\\
\hline\hline
$-0.4$ & $0.0$ & $1.3746$ & $2.0358$ & $3.4282$ & $4.1436$ & $0.3081$\\
\hline
$-0.4$ & $0.4$ & $1.3099$ & $1.9479$ & $3.3115$ & $3.9380$ & $0.3132$\\
\hline
$-0.4$ & $0.8$ & $1.0648$ & $1.6225$ & $2.8904$ & $3.2096$ & $0.3265$\\
\hline
$-0.2$ & $0.0$ & $1.6944$ & $2.5266$ & $4.3225$ & $5.0880$ & $0.2371$\\
\hline
$-0.2$ & $0.4$ & $1.6198$ & $2.4264$ & $4.1912$ & $4.8595$ & $0.2399$\\
\hline
$-0.2$ & $0.8$ & $1.3367$ & $2.0579$ & $3.7225$ & $4.0567$ & $0.2453$\\
\hline
$0.0$ & $0.0$ & $2.0000$ & $3.0000$ & $5.1962$ & $6.0000$ & $0.1925$\\
\hline
$0.0$ & $0.4$ & $1.9165$ & $2.8892$ & $5.0530$ & $5.7528$ & $0.1941$\\
\hline
$0.0$ & $0.8$ & $1.6000$ & $2.4849$ & $4.5460$ & $4.8908$ & $0.1959$\\
\hline
$0.3$ & $0.0$ & $2.4402$ & $3.6885$ & $6.4823$ & $7.3298$ & $0.1498$\\
\hline
$0.3$ & $0.4$ & $2.3450$ & $3.5645$ & $6.3245$ & $7.0597$ & $0.1505$\\
\hline
$0.3$ & $0.8$ & $1.9841$ & $3.1155$ & $5.7718$ & $6.1275$ & $0.1499$\\
\hline
$0.6$ & $0.0$ & $2.8649$ & $4.3596$ & $7.7493$ & $8.6297$ & $0.1224$\\
\hline
$0.6$ & $0.4$ & $2.7593$ & $4.2243$ & $7.5795$ & $8.3411$ & $0.1226$\\
\hline
$0.6$ & $0.8$ & $2.3589$ & $3.7387$ & $6.9907$ & $7.3539$ & $0.1211$\\
\hline
$1.0$ & $0.0$ & $3.4142$ & $5.2361$ & $9.4192$ & $10.3329$ & $0.0981$\\
\hline
$1.0$ & $0.4$ & $3.2961$ & $5.0881$ & $9.2364$ & $10.0242$ & $0.0981$\\
\hline
$1.0$ & $0.8$ & $2.8485$ & $4.5620$ & $8.6092$ & $8.9793$ & $0.0961$\\
\hline
$1.5$ & $0.0$ & $4.0811$ & $6.3117$ & $11.4862$ & $12.4305$ & $0.0784$\\
\hline
$1.5$ & $0.4$ & $3.9491$ & $6.1505$ & $11.2903$ & $12.1018$ & $0.0783$\\
\hline
$1.5$ & $0.8$ & $3.4487$ & $5.5837$ & $10.6258$ & $11.0018$ & $0.0763$\\
\hline
$2.0$ & $0.0$ & $4.7321$ & $7.3723$ & $13.5383$ & $14.5054$ & $0.0652$\\
\hline
$2.0$ & $0.4$ & $4.5875$ & $7.2000$ & $13.3318$ & $14.1608$ & $0.0650$\\
\hline
$2.0$ & $0.8$ & $4.0392$ & $6.6000$ & $12.6380$ & $13.0180$ & $0.0631$\\
\hline
\end{tabular}
\caption{Null and marginally stable structure of the charged STVG spacetime. Columns list the
coupling, the reduced charge $q=Q/M$, the event horizon of Eq.~(\ref{eq:horizons}), the photon
sphere of Eq.~(\ref{eq:rph}), the shadow radius of Eq.~(\ref{eq:bcrit}), the innermost stable
circular orbit obtained from Eq.~(\ref{eq:iscocubic}) and the Lyapunov exponent of
Eq.~(\ref{eq:lyap}). The row $\alpha=q=0$ returns the Schwarzschild values $2$, $3$, $3\sqrt3$, $6$
and $1/3\sqrt3$ exactly.}
\label{tab:null}
\end{table}

\begin{figure}[ht!]
\centering
\includegraphics[width=0.86\textwidth]{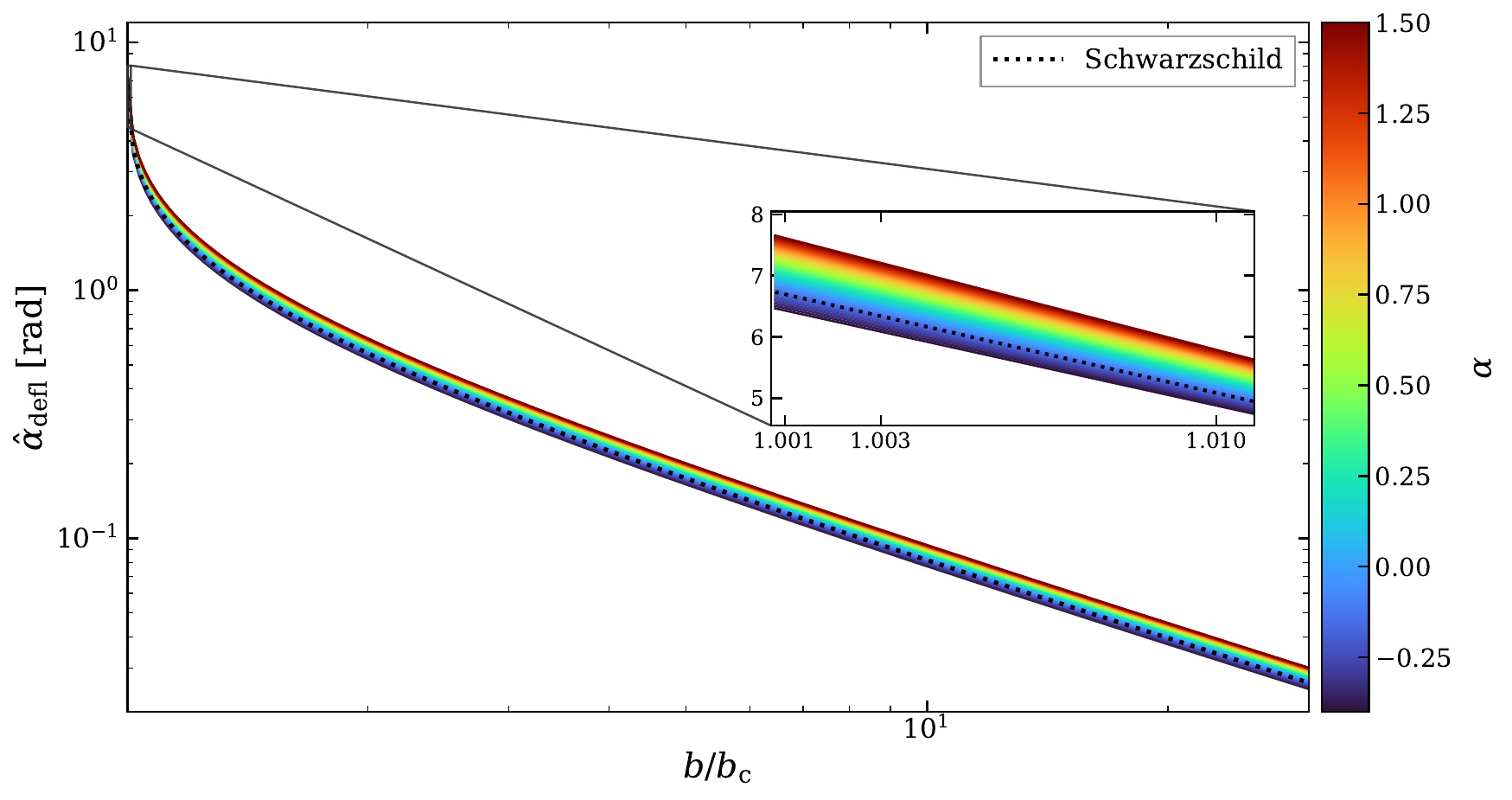}
\caption{Deflection angle of Eq.~(\ref{eq:defl}) against impact parameter in units of the critical
one, for $q=0.4$ and a family of couplings $-0.4\le\alpha\le1.5$ encoded by the ribbon color, with
the dotted black curve giving the Schwarzschild result. The inset magnifies the interval
$1.0008\le b/b_{\rm c}\le1.010$, where the family is most tightly bunched, and the connecting guide
lines mark the region it magnifies. The logarithmic divergence as $b\to b_{\rm c}$ is common to all
members of the family, and the $\alpha$ dependence enters only through the value of $b_{\rm c}$
itself, so the curves nearly superimpose once the abscissa is scaled.}
\label{fig:deflect}
\end{figure}

Figure \ref{fig:deflect} makes a point that will matter when the QPO bounds are interpreted. Once the impact parameter is measured in units of $b_{\rm c}$, the deflection curves for
different couplings collapse almost onto one another. The STVG modification therefore rescales the
lensing problem rather than reshaping it, which is again a consequence of Eq.~(\ref{eq:map}): the
dimensionless deflection function depends on the single ratio $\Qeff/\Meff$, and this ratio varies
only weakly across the family shown. Table \ref{tab:null} gives the numerical values of all five
characteristic quantities over a grid in $\alpha$ and $q$, and shows the same competition seen in
Fig.~\ref{fig:lapse}: the shadow radius grows along $\alpha$ and shrinks along $q$. A degeneracy
line exists along which the shadow is unchanged from its Schwarzschild value, and any object whose
shadow has been measured but whose QPO fit prefers nonzero $\alpha$ must lie close to it.

\section{Circular timelike geodesics and the innermost stable orbit}\label{isec4}

We now assemble the ingredients that the RP model actually uses. The program is standard: write
the Lagrangian for a neutral massive test particle, identify the conserved quantities, impose the
circular orbit conditions, and read off the orbital frequency together with the specific energy and
angular momentum. What is specific to the present spacetime is the algebraic form of the results and
the way the photon-sphere polynomial of Eq.~(\ref{eq:phquad}) reappears in the normalization, which
we use as a check at each stage.

The Lagrangian of a test particle of mass $m$ is
\begin{equation}
\mathcal{L}=\tfrac{1}{2}m\,g_{\mu\nu}\dot x^{\mu}\dot x^{\nu},
\label{eq:lag}
\end{equation}
with $\dot x^{\mu}=\dd x^{\mu}/\dd\tau$ the four-velocity along the worldline $x^{\mu}(\tau)$. The
metric of Eq.~(\ref{eq:metric}) is static and axially symmetric, so the specific energy and the
specific angular momentum
\begin{equation}
g_{tt}\dot t=-E,\qquad g_{\varphi\varphi}\dot\varphi=L
\label{eq:conserved}
\end{equation}
are conserved. Since $m\neq0$ the equations of motion read
\begin{equation}
\dot t=-\frac{E}{g_{tt}},\qquad \dot\varphi=\frac{L}{g_{\varphi\varphi}},
\label{eq:eom}
\end{equation}
\begin{equation}
g_{rr}\dot r^{2}+g_{\theta\theta}\dot\theta^{2}=V_{\rm eff},
\label{eq:radial}
\end{equation}
with the effective potential
\begin{equation}
V_{\rm eff}(r)=-\left(1+\frac{E^{2}g_{\varphi\varphi}+L^{2}g_{tt}}{g_{tt}g_{\varphi\varphi}}\right).
\label{eq:veff}
\end{equation}
For circular motion in the equatorial plane, $\dot r=\dot\theta=0$, the orbital parameters follow as
\begin{equation}
\Omega_{\varphi}=\sqrt{-\frac{\partial_{r}g_{tt}}{\partial_{r}g_{\varphi\varphi}}},\qquad
\dot t=u^{t}=\frac{1}{\sqrt{-g_{tt}-g_{\varphi\varphi}\Omega_{\varphi}^{2}}},
\label{eq:omut}
\end{equation}
\begin{equation}
E=-\frac{g_{tt}}{\sqrt{-g_{tt}-g_{\varphi\varphi}\Omega_{\varphi}^{2}}},\qquad
L=\frac{g_{\varphi\varphi}\Omega_{\varphi}}{\sqrt{-g_{tt}-g_{\varphi\varphi}\Omega_{\varphi}^{2}}} .
\label{eq:EL}
\end{equation}
Inserting Eq.~(\ref{eq:lapse}) and using $\Omega_{\varphi}^{2}=f'(r)/2r$ gives the closed forms
\begin{equation}
\Omega\equiv\Omega_{\varphi}=\frac{\sqrt{(1+\alpha)\left[M(r-\alpha M)-Q^{2}\right]}}{r^{2}},
\label{eq:omphi}
\end{equation}
\begin{equation}
E=\frac{r^{2}+(1+\alpha)\left(\alpha M^{2}-2Mr+Q^{2}\right)}{r\sqrt{r^{2}+(1+\alpha)\left(2\alpha M^{2}-3Mr+2Q^{2}\right)}},
\label{eq:Ecirc}
\end{equation}
\begin{equation}
L=\frac{r\sqrt{(1+\alpha)\left[M(r-\alpha M)-Q^{2}\right]}}{\sqrt{r^{2}+(1+\alpha)\left(2\alpha M^{2}-3Mr+2Q^{2}\right)}} .
\label{eq:Lcirc}
\end{equation}
The radicand shared by Eqs.~(\ref{eq:Ecirc}) and (\ref{eq:Lcirc}) is
$r^{2}-3(1+\alpha)Mr+2(1+\alpha)(\alpha M^{2}+Q^{2})$, which is exactly the photon-sphere polynomial
of Eq.~(\ref{eq:phquad}). Both $E$ and $L$ therefore diverge as $r\to\rps^{+}$, the expected
behavior since a circular timelike orbit becomes null there, and this identity confirms that
Eqs.~(\ref{eq:rph}), (\ref{eq:Ecirc}) and (\ref{eq:Lcirc}) are mutually consistent. Circular orbits
exist only for $r>\rps$, and Eq.~(\ref{eq:omphi}) requires in addition $r>C/M$, a weaker condition in
the parameter range of interest.

In terms of the reduced variables of Eq.~(\ref{eq:shorthand}) the orbital frequency reads
$M^{2}\Omega_{\varphi}^{2}=A(x-c)/x^{4}$, which for $A=1$, $c=0$ collapses to the Keplerian
$M^{2}\Omega^{2}=1/x^{3}$. The specific angular momentum has a minimum at the marginally stable
orbit, and the condition $\dd L/\dd r=0$, equivalently $\dd E/\dd r=0$, yields after simplification
the cubic
\begin{equation}
x^{3}-6Ax^{2}+9Acx-4Ac^{2}=0 ,
\label{eq:iscocubic}
\end{equation}
whose largest real root outside the horizon is $x_{\rm ISCO}=r_{\rm ISCO}/M$. For $A=1$, $c=q^{2}$
this is the familiar RN condition $x^{3}-6x^{2}+9q^{2}x-4q^{4}=0$, and for $c=0$ it gives
$x_{\rm ISCO}=6$. The cubic admits the closed-form root
\begin{equation}
\frac{r_{\rm ISCO}}{M}=\frac{B_{1}^{2}+A_{1}+(1+\alpha)\left(2B_{1}-3q^{2}\right)}{B_{1}},
\label{eq:iscoclosed}
\end{equation}
with
\begin{equation}
B_{1}=\left[B_{2}+A_{2}-A_{3}q^{2}+2(1+\alpha)q^{4}\right]^{1/3},\qquad
B_{2}=(1+\alpha)\left(\alpha+q^{2}\right)\sqrt{\left(1-q^{2}\right)\left(5+\alpha-4q^{2}\right)},
\label{eq:B1B2}
\end{equation}
\begin{equation}
A_{1}=4+5\alpha+\alpha^{2},\qquad
A_{2}=8+15\alpha+8\alpha^{2}+\alpha^{3},\qquad
A_{3}=9+14\alpha+5\alpha^{2}.
\label{eq:A123}
\end{equation}
We have verified that Eq.~(\ref{eq:iscoclosed}) satisfies Eq.~(\ref{eq:iscocubic}) identically: at
the representative points $(\alpha,q)=(0,0)$, $(0,0.5)$, $(0.5,0)$, $(0.3,0.4)$, $(1,0.2)$ and
$(-0.3,0.1)$ the residual of the cubic evaluated on Eq.~(\ref{eq:iscoclosed}) is below
$6\times10^{-14}$, and the returned radii $6$, $5.6066$, $8.1990$, $7.0597$, $10.2572$ and $4.6076$
coincide with the numerically extracted largest root in every case. The first of these is the
Schwarzschild value and the second the RN value at $q=0.5$. The ISCO fixes the inner edge of the
accretion disc and hence the emitting area available to the flow, and its dependence on the spacetime
parameters has been used in the same spirit for rotating and strange-star exteriors
\citep{Stergioulas:1999,Zdunik:2000,Luk:2018,Boshkayev:2020} and in searches for its imprint in the
kilohertz timing data themselves \citep{Barret:2007}.

\begin{figure}[t!]
\centering
\includegraphics[width=\textwidth]{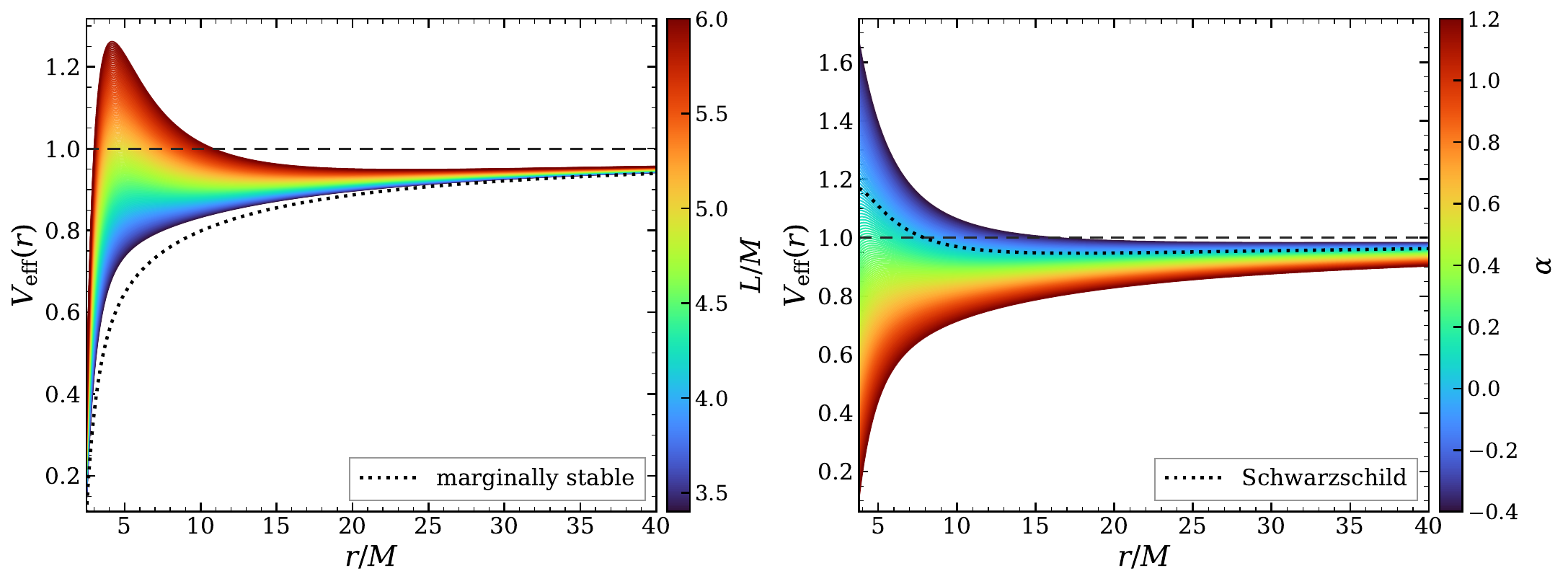}
\vspace{-0.3cm}\\
\caption{Effective potential $V_{\rm eff}=f(r)\left(1+L^{2}/r^{2}\right)$ for neutral massive test
particles. The left panel fixes $(\alpha,q)=(0.3,0.4)$ and sweeps the specific angular momentum over
$3.4\le L/M\le6.0$; the dotted black curve is the marginally stable member of the family, for which
the local maximum and minimum merge at the ISCO. The right panel fixes $L/M=4.6$ and $q=0.4$ and
sweeps the coupling over $-0.4\le\alpha\le1.2$, with the dotted black curve giving the Schwarzschild
potential at the same angular momentum. The horizontal dashed line at $V_{\rm eff}=1$ separates bound
from unbound motion.}
\label{fig:veff}
\end{figure}

Figure \ref{fig:veff} shows how the potential barrier responds to the two parameters. Raising the
coupling deepens the well and moves both extrema outward, so a particle of given angular momentum
that would orbit stably in the Schwarzschild geometry can be marginally stable or plunging in the
STVG one. Raising the charge acts in the opposite direction and stiffens the centrifugal barrier at
small radius. The left panel makes the ISCO construction explicit: as $L$ decreases the maximum and
the minimum approach each other, and their merger defines the marginally stable orbit given in
closed form by Eq.~(\ref{eq:iscoclosed}).

\begin{figure}[t!]
\centering
\includegraphics[width=\textwidth]{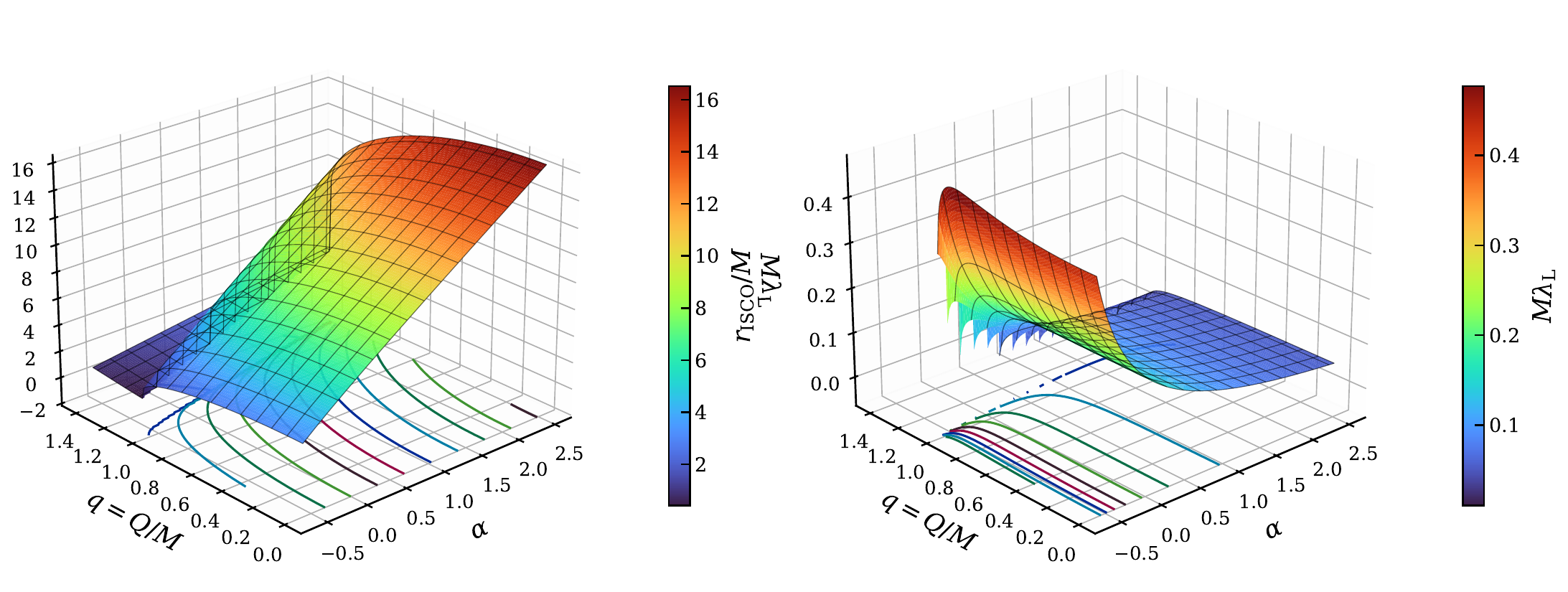}
\vspace{-0.4cm}\\
\caption{Three-dimensional surfaces over the $(\alpha,q)$ plane. Left: ISCO radius in units of $M$,
obtained as the outermost root of Eq.~(\ref{eq:iscocubic}). Right: Lyapunov exponent
$M\lambda_{\rm L}$ of the circular null orbit from Eq.~(\ref{eq:lyap}). Contours of each surface
are projected onto the floor of its box. The Schwarzschild values $r_{\rm ISCO}=6M$ and
$M\lambda_{\rm L}=1/3\sqrt{3}$ are recovered at $\alpha=q=0$, and the two surfaces slope in
opposite senses because a larger marginally stable orbit accompanies a longer instability
timescale.}
\label{fig:maps}
\end{figure}

Figure \ref{fig:maps} presents the ISCO radius and the Lyapunov exponent as surfaces over the full
parameter plane. The two run in opposite senses, as they must: a larger ISCO goes with a larger
photon sphere and hence a longer instability timescale, so $\lambda_{\rm L}$ decreases where
$r_{\rm ISCO}$ increases. The projected contours of both quantities are close to straight lines of
negative slope in the $(\alpha,q)$ plane over the region of interest, which is the geometric expression of the
degeneracy of Eq.~(\ref{eq:map}) and a first indication that fits will not localize $\alpha$ and $Q$
independently.

\section{Epicyclic frequencies and the relativistic precession model}\label{isec5}

The observable content of the model lies in the two epicyclic frequencies, and this section derives
them, fixes the radial one by explicit limit checks, and assembles the prediction
that will be confronted with the data. The Schwarzschild limit alone does not fix the
$r^{-5}$ and $r^{-6}$ coefficients, so both limits are quoted. We give the derivation in
enough detail that the result can be verified independently, and we list the three separate checks
that fix it.

Perturbing a circular orbit by $r\to r_{0}+\delta r$ and $\theta\to\pi/2+\delta\theta$ and
linearizing Eq.~(\ref{eq:radial}) gives harmonic motion,
\begin{equation}
\frac{\dd^{2}\delta r}{\dd t^{2}}+\Omega_{r}^{2}\,\delta r=0,\qquad
\frac{\dd^{2}\delta\theta}{\dd t^{2}}+\Omega_{\theta}^{2}\,\delta\theta=0,
\label{eq:harm}
\end{equation}
with the squared epicyclic frequencies measured by a distant observer,
\begin{equation}
\Omega_{r}^{2}=-\frac{1}{2g_{rr}\left(u^{t}\right)^{2}}\frac{\partial^{2}V_{\rm eff}}{\partial r^{2}}\bigg|_{\theta=\pi/2},
\qquad
\Omega_{\theta}^{2}=-\frac{1}{2g_{\theta\theta}\left(u^{t}\right)^{2}}\frac{\partial^{2}V_{\rm eff}}{\partial\theta^{2}}\bigg|_{\theta=\pi/2} .
\label{eq:epidef}
\end{equation}
An equivalent and computationally cleaner route writes the equatorial radial equation as
$\dot r^{2}=E^{2}-f(r)\left(1+L^{2}/r^{2}\right)\equiv\mathcal{R}(r)$. A circular orbit at $r_{0}$
satisfies $\mathcal{R}(r_{0})=\mathcal{R}'(r_{0})=0$, and the proper-time oscillation frequency is
$\omega_{r}^{2}=-\tfrac{1}{2}\mathcal{R}''(r_{0})$, which converts to coordinate time on division by
$\left(u^{t}\right)^{2}$. Carrying this out with Eqs.~(\ref{eq:Ecirc}) and (\ref{eq:Lcirc}) inserted
and simplifying yields
\begin{equation}
\Omega_{r}^{2}=\frac{(1+\alpha)M}{r^{3}}-\frac{6(1+\alpha)^{2}M^{2}}{r^{4}}
+\frac{9(1+\alpha)^{2}M\left(\alpha M^{2}+Q^{2}\right)}{r^{5}}
-\frac{4(1+\alpha)^{2}\left(\alpha M^{2}+Q^{2}\right)^{2}}{r^{6}} ,
\label{eq:omr2}
\end{equation}
or compactly, in the notation of Eq.~(\ref{eq:shorthand}),
\begin{equation}
\Omega_{r}^{2}=\frac{A\,M\,r^{3}-6A^{2}M^{2}r^{2}+9A^{2}M C\,r-4A^{2}C^{2}}{r^{6}},
\qquad
M^{2}\Omega_{r}^{2}=\frac{A\left(x^{3}-6Ax^{2}+9Acx-4Ac^{2}\right)}{x^{6}} .
\label{eq:omr2compact}
\end{equation}
Three checks fix this expression. First, the Schwarzschild limit $\alpha=Q=0$ gives
$\Omega_{r}^{2}=(M/r^{3})(1-6M/r)$. Second, the RN limit $\alpha=0$ gives
\begin{equation}
\Omega_{r}^{2}\Big|_{\alpha=0}=\frac{1}{r^{4}}\left(Mr-6M^{2}+\frac{9MQ^{2}}{r}-\frac{4Q^{4}}{r^{2}}\right),
\label{eq:rnlimit}
\end{equation}
which is the standard result for the charged case. Third, the vanishing of
Eq.~(\ref{eq:omr2compact}) reproduces the ISCO cubic of Eq.~(\ref{eq:iscocubic}) exactly, as it must,
since the marginally stable orbit is where the radial oscillation frequency goes to zero. We note that the closed-form ISCO of Eq.~(\ref{eq:iscoclosed}) is the root of the cubic
generated by Eq.~(\ref{eq:omr2compact}), so the ISCO and the radial frequency corroborate each
other. In terms of the effective parameters, Eq.~(\ref{eq:omr2compact}) is again the RN
expression with $M\to\Meff$ and $Q^{2}\to\Qeff^{2}$, which is consistent with the isometry of Eq.~(\ref{eq:map}).

The vertical sector is simpler. The line element of Eq.~(\ref{eq:metric}) is spherically symmetric,
so motion out of the equatorial plane is not coupled to the radial motion and the vertical epicyclic
frequency coincides with the orbital one,
\begin{equation}
\Omega_{\theta}=\Omega_{\varphi}=\sqrt{\frac{(1+\alpha)M}{r^{3}}-\frac{(1+\alpha)\left(\alpha M^{2}+Q^{2}\right)}{r^{4}}} .
\label{eq:omth}
\end{equation}
The absence of nodal precession is a signature of staticity: any rotation of the central object, and
in particular the frame dragging carried by the Hartle--Thorne exterior used in
Ref.~\citep{Boshkayev:2026}, would split $\Omega_{\theta}$ from $\Omega_{\varphi}$ and introduce a
third frequency. That a static model can nevertheless describe the twin-peak data as well as it does
is one of the results of Section \ref{isec7}, and it also indicates the limitation of the present
treatment, since the deviation attributed to $\alpha$ and $Q$ could in part be absorbing rotational
effects.

\begin{figure}[ht!]
\centering
\includegraphics[width=\textwidth]{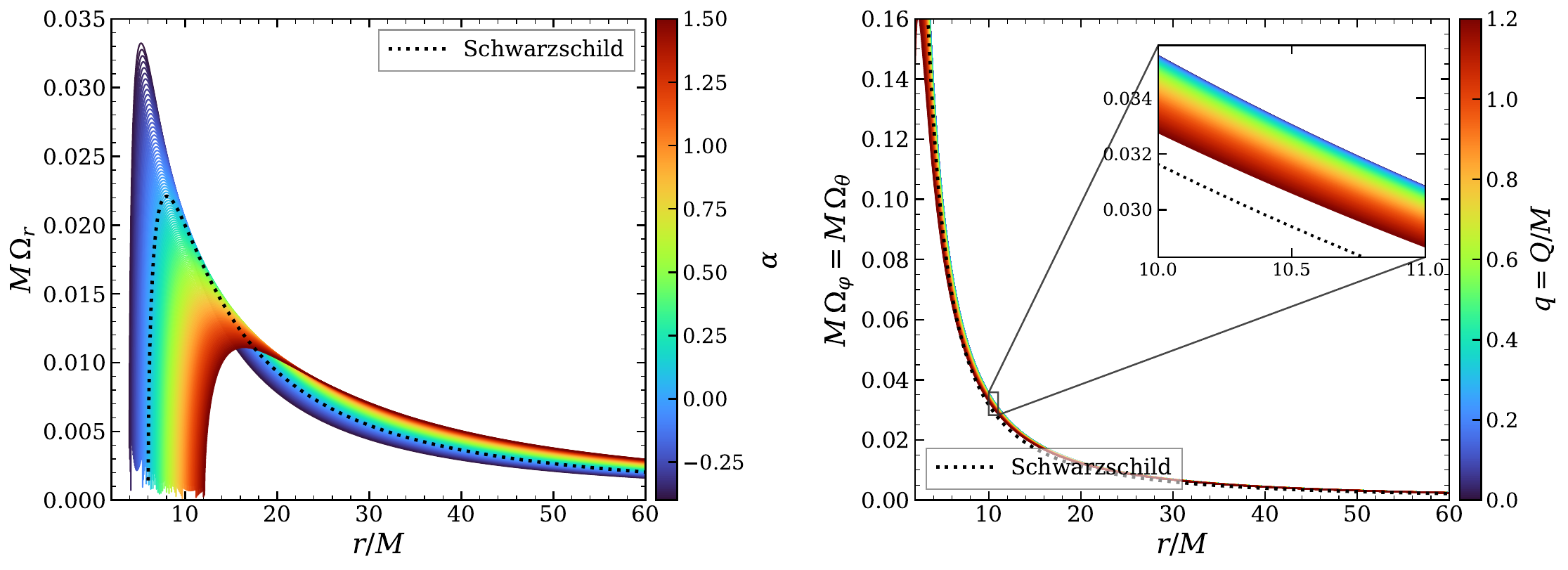}
\vspace{-0.3cm}\\
\caption{Fundamental frequencies of neutral test particles on circular equatorial orbits. Left: the
radial epicyclic frequency of Eq.~(\ref{eq:omr2compact}) at $q=0.4$ for couplings
$-0.4\le\alpha\le1.5$ encoded by the ribbon color, with the dotted black curve giving the
Schwarzschild profile. Each curve vanishes at the corresponding ISCO and peaks a little outside it.
The vertical range is truncated at $M\Omega_{r}=0.035$. Right: the orbital frequency of
Eq.~(\ref{eq:omphi}), which by Eq.~(\ref{eq:omth}) is also the vertical epicyclic frequency, at
$\alpha=0.3$ for reduced charges $0\le q\le1.2$; the inset magnifies $10\le r/M\le11$, where the
family is most tightly bunched, with guide lines marking the region it magnifies. Raising $\alpha$
lowers and broadens the radial profile while pushing its zero outward; raising $q$ raises
$\Omega_{\varphi}$ at small radius.\\}
\label{fig:freqs}
\end{figure}

Figure \ref{fig:freqs} shows both frequencies as functions of radius. The radial frequency has the
characteristic shape imposed by Eq.~(\ref{eq:omr2compact}): it vanishes at the ISCO, rises to a
maximum a few gravitational radii further out, and then falls off. The location and height of that
maximum are what the QPO data are sensitive to, since the lower peak in the RP model is the
difference between the orbital and radial frequencies and is therefore largest where $\Omega_{r}$ is
smallest relative to $\Omega_{\varphi}$.

The RP model converts these angular frequencies into observed frequencies. Writing
$f_{\varphi}=\Omega_{\varphi}/2\pi$ for the Keplerian frequency and $f_{r}=\Omega_{r}/2\pi$ for the
radial epicyclic frequency of Keplerian motion, the model identifies the upper peak with the
Keplerian frequency and the lower peak with the periastron precession frequency,
\begin{equation}
f_{\rm U}=f_{\varphi},\qquad f_{\rm L}=f_{\varphi}-f_{r} .
\label{eq:rp}
\end{equation}
Eliminating the orbital radius between the two relations gives a one-parameter curve
$f_{\rm L}(f_{\rm U})$ for each choice of $\{M,\alpha,Q\}$, and it is this curve that is compared
with the observed frequency pairs. The curve starts at the origin, since both frequencies vanish at
large radius, rises with a slope that approaches unity in the Newtonian regime, and turns over as the
orbital radius approaches the ISCO, where $f_{r}\to0$ and hence $f_{\rm L}\to f_{\rm U}$. The
position of that turnover is the discriminating feature: it occurs at
$f_{\rm U}=f_{\varphi}(r_{\rm ISCO})$, which through Eq.~(\ref{eq:iscocubic}) depends on all three
parameters.

\begin{figure}[ht!]
\centering
\includegraphics[width=\textwidth]{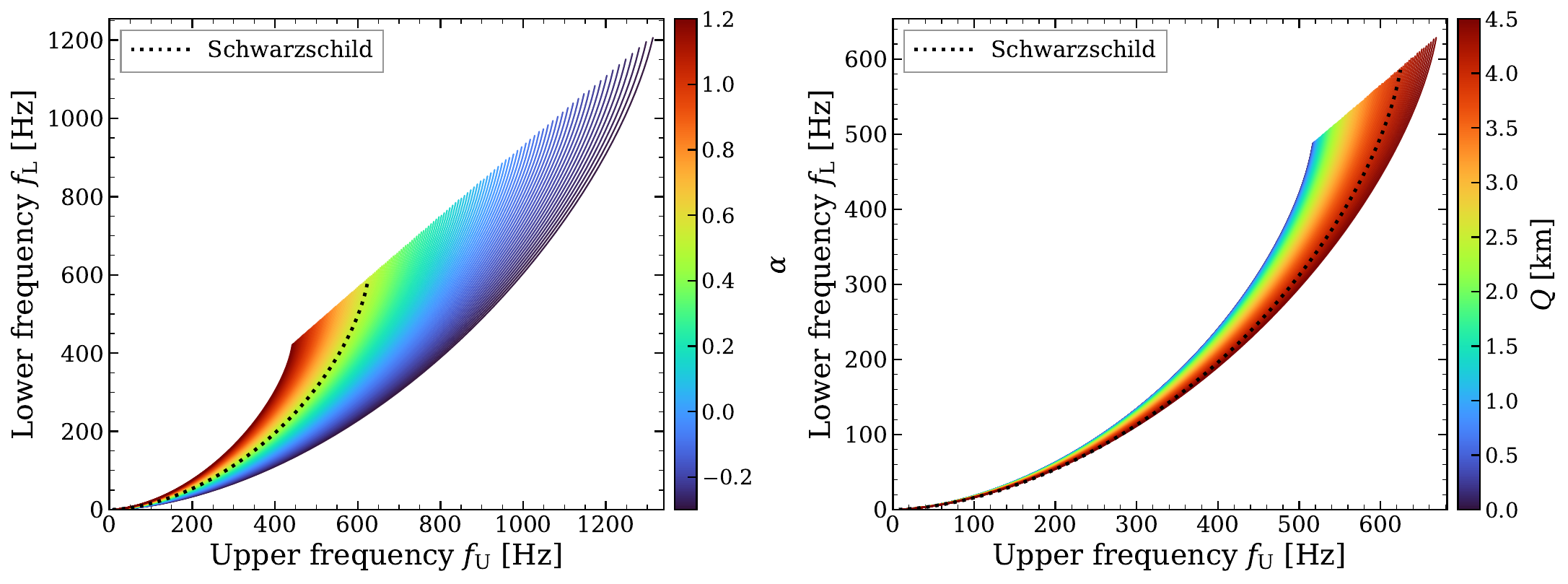}
\vspace{-0.3cm}\\
\caption{RP prediction of Eq.~(\ref{eq:rp}) for a central object of
$M=3.5\,\Msun$. Left: fixed $Q=5$ km with the coupling swept over $-0.3\le\alpha\le1.2$. Right:
fixed $\alpha=0.3$ with the charge swept over $0\le Q\le4.5$ km. The dotted black curve in both
panels is the Schwarzschild prediction at the same mass. Each curve is drawn only on the branch that
lies outside the marginally stable orbit, so that the pair always refers to a stable circular orbit,
and it terminates where $f_{r}$ vanishes at the ISCO. Increasing $\alpha$ shifts the whole relation
to lower frequencies and flattens it, while increasing $Q$ shifts it to higher frequencies; the two effects partly cancel, which is the origin of the degeneracy discussed in Section \ref{isec7}.\\}
\label{fig:rp}
\end{figure}

Figure \ref{fig:rp} displays the families of theoretical curves generated by varying each parameter
in turn. The two panels show opposite trends, and comparing them makes the compensation explicit: a
curve computed with larger $\alpha$ can be brought back onto a curve computed with smaller $\alpha$
by increasing $Q$, and by Eq.~(\ref{eq:map}) the compensation is exact whenever $\Meff$ and $\Qeff$
are held fixed. The flattening at high $f_{\rm U}$ produced by increasing $\alpha$ is the feature
that the neutron star data will be seen to prefer.

\subsection{QPO Models and Orbital Dynamics in Black Hole Spacetimes}

This subsection explores the potential frequencies of twin-peak QPOs in static charged solutions of STV Gravity. We compare these findings with the results for standard Schwarzschild black holes across various QPO models. The upper $\left(f_{U}\right)$ and lower $\left(f_{L}\right)$ QPO frequencies are presented in terms of $f_r$, $f_\theta$, and $f_\phi$, as formulated according to each specific QPO model under consideration \cite{QPO4}. For instance, we adopt the following commonly used models.

\subsubsection{Warped Disk (WD) model}
The WD model explains high-frequency QPOs by examining a non-standard, warped geometry of the accretion disk \cite{QPO13}. Within this framework, nonlinear resonances arise between the warped structure of the disk and its intrinsic oscillation modes. These resonances involve interactions in both horizontal and vertical directions. Horizontal resonances can excite both g-mode and p-mode oscillations, while vertical resonances typically excite only g-modes. Indeed, the emergence of these resonances is due to the non-monotonic variation of the radial epicyclic frequency in relation to the radial coordinate $r$. In this model, the observed high-frequency QPOs correspond to the relationships $f_U = 2f_\phi - f_r$ and $f_L = 2(f_\phi - f_r)$.

\subsubsection{ER2, ER3, and ER4 Models}
The epicyclic resonance (ER) model provides a relativistic framework for interpreting high-frequency QPOs as a result of nonlinear resonances among the fundamental oscillation modes of accreting matter near compact objects. In this model, QPOs emerge from resonances between the radial and vertical epicyclic frequencies of particles that are orbiting along slightly perturbed geodesic trajectories within a relativistic accretion disk. In this work, we concentrate on three representative cases of the epicyclic resonance (ER) model: ER2, ER3, and ER4. Each case is defined by unique combinations of the orbital and epicyclic frequencies. The QPO frequencies for these models are 
defined as follows \cite{QPO1}:
\begin{itemize}
\item For ER2, the upper and lower QPO frequencies are given by 
$f_U = 2f_\theta - f_r$ and $f_L = f_r$.
\item For ER3, they are expressed as $f_U = f_\theta + f_r$ and 
$f_L = f_\theta$.
\item For ER4, the relations take the form: 
$f_U = f_\theta + f_r$ and $f_L = f_\theta - f_r$.
\end{itemize}

\begin{figure}[ht!]
\centering
\includegraphics[width=0.48\textwidth,clip]{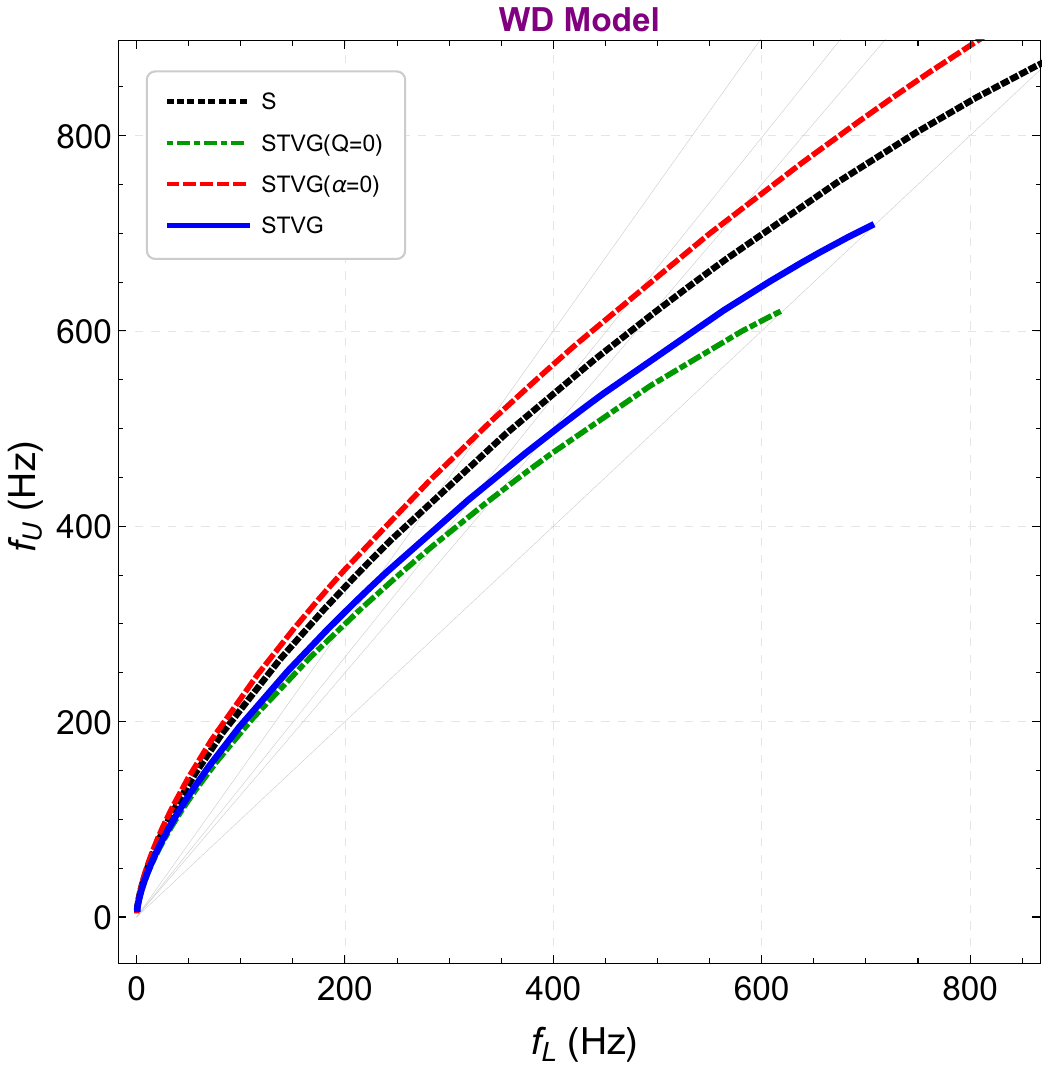}\hspace{10.8pt}%
\includegraphics[width=0.48\textwidth,clip]{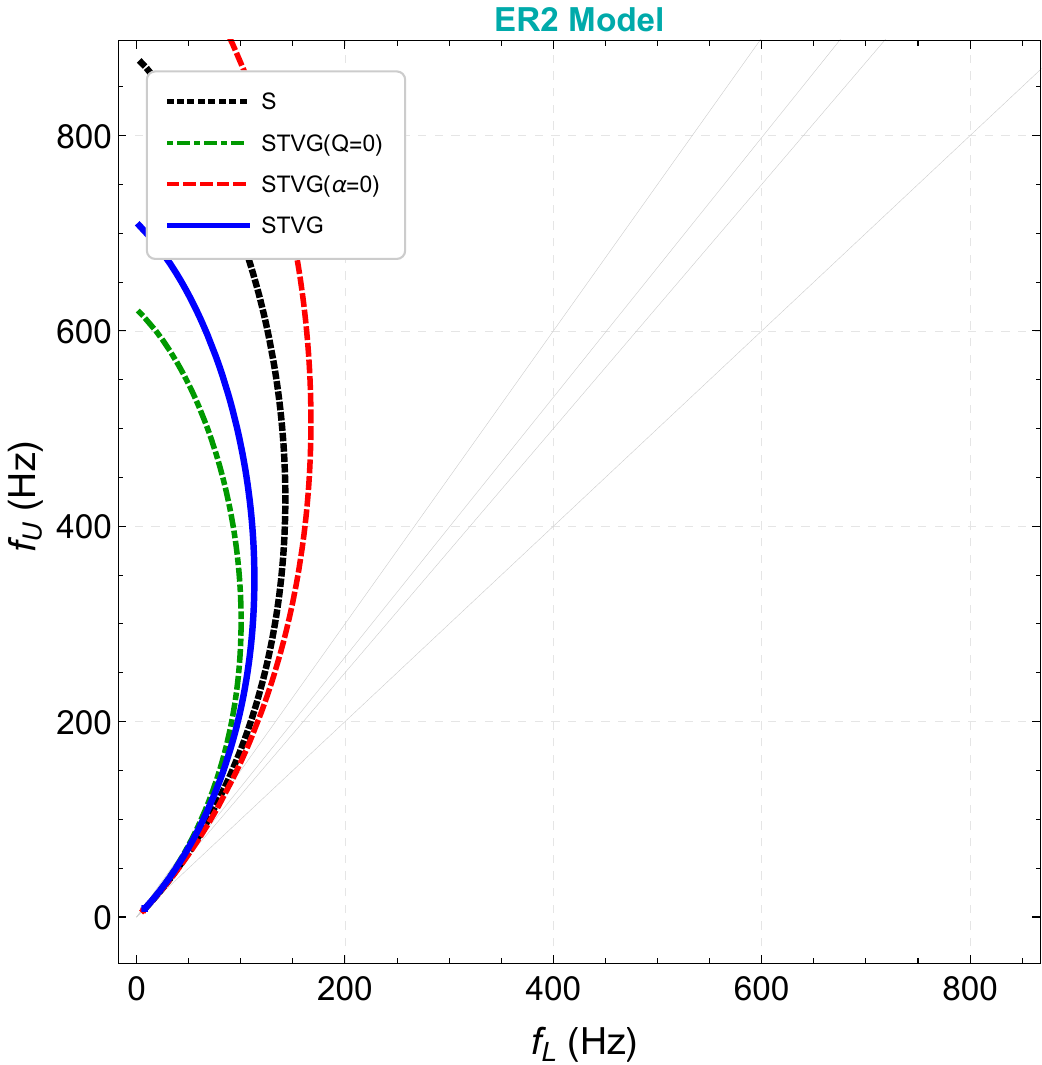}\\[4pt]
\includegraphics[width=0.48\textwidth,clip]{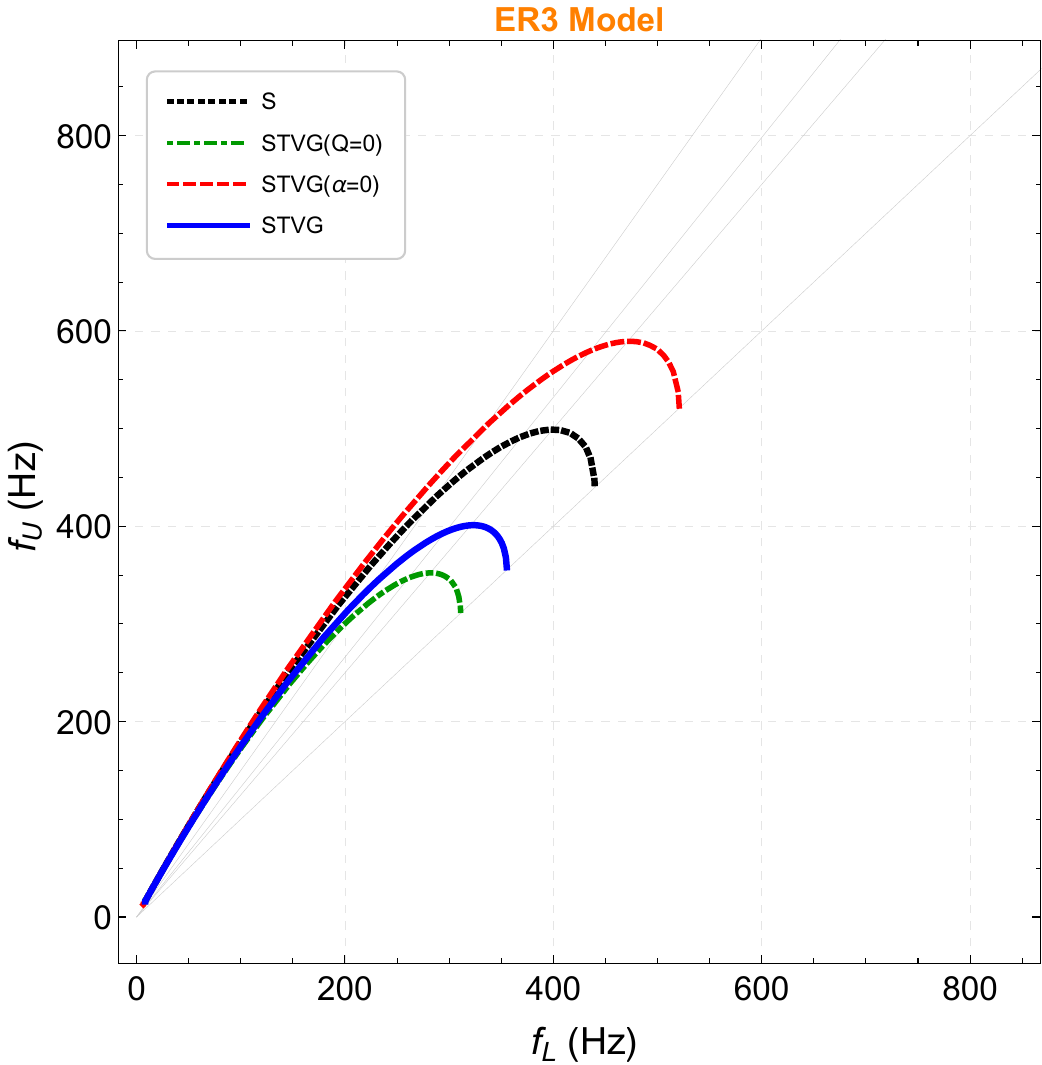}\hspace{10.8pt}%
\includegraphics[width=0.48\textwidth,clip]{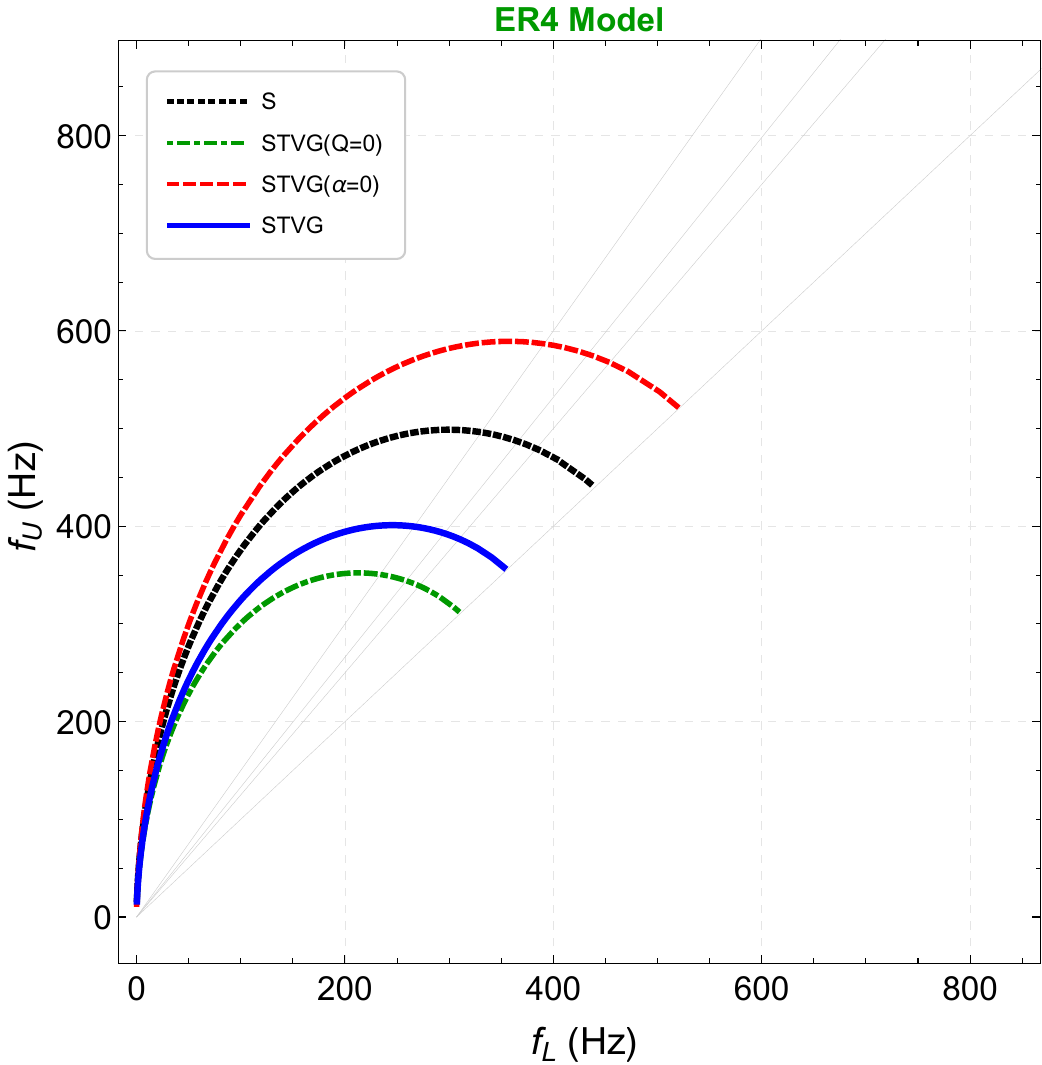}

\vspace{5pt}

\caption{The relations between the two twin-peak QPOs frequencies, namely upper ($f_U$) and lower ($f_L$) frequencies for different configurations of STVG parameters predicted by WD, ER2, ER3 and ER4. Here we have chosen $\bar M=5.0 M_\odot$. The curves represent the STVG full model with $\alpha=0.6, Q=0.7$ (solid blue curve), pure charge $\alpha=0, Q=0.7$ (dashed red curve), pure deformation $\alpha=0.6, Q=0$ (dot-dashed green curve) and Schwarzschild baseline $\alpha=0, Q=0$ (dotted black curve).\\}
\label{QPOS models}
\end{figure}

Figure \ref{QPOS models} illustrates the correlations between the upper ($f_U$) and lower ($f_L$) frequencies of twin-peak quasi-periodic oscillations (QPOs) within the context of STVG spacetime black holes. This process utilizes the WD, ER2, ER3, and ER4 models while varying the STVG parameters ($\alpha$ and $Q$), all at a constant black hole mass of $\bar{M} = 5.0 M_{\odot}$. The plot features light ray reference lines that illustrate the rational frequency ratios of 3:2, 4:3, 5:4, and 1:1, demonstrating the proportional relationship between the upper and lower frequencies. The diagonal line corresponding to the 1:1 frequency ratio is referred to as the QPO graveyard. When a QPO trajectory reaches this limit, it indicates that the upper and lower frequencies have merged into a single oscillation, thereby eliminating the twin-peak structure behavior.  The introduction of the STVG deformation parameter ($\alpha = 0.6$) and electric charge ($Q = 0.7$) clearly illustrates a progressive deviation of the predicted frequency correlations from the standard Schwarzschild reference curve (S, dotted black line) across all four panels. This noticeable shift emphasizes the significant impact of STVG modified gravity on observable QPO characteristics and the dynamics of orbital motion within the strong-field regime.

\section{Statistical framework}\label{isec6}

The comparison between the theoretical curve and the data is carried out with a likelihood analysis
followed by a model selection step, and we describe both here before presenting results. Two
decisions deserve emphasis. We fit $f_{\rm L}$ as a function of the measured $f_{\rm U}$ rather than
fitting both frequencies against an unobserved orbital radius, which avoids introducing one nuisance
parameter per data point in the neutron star sample. And we compare models with the deviance
information criterion rather than with the likelihood alone, so that the extra freedom carried by the
STVG parameters is penalized in a controlled way.

For each source we take $N$ twin kilohertz frequency pairs with their uncertainties, written
$\left(f_{{\rm L},k}\pm\sigma_{{\rm L},k},\,f_{{\rm U},k}\pm\sigma_{{\rm U},k}\right)$, and identify
the best fit as the parameter set maximizing the log-likelihood
\begin{equation}
\ln \mathcal{L}=-\frac{1}{2}\sum_{k=1}^{N}\left\{
\frac{\left[f_{{\rm L},k}-f_{\rm L}\!\left(h,f_{{\rm U},k}\right)\right]^{2}}{\sigma_{{\rm L},k}^{2}}
+\ln\!\left(2\pi\sigma_{{\rm L},k}^{2}\right)\right\},
\label{eq:loglike}
\end{equation}
where $f_{\rm L}$ is evaluated from Eq.~(\ref{eq:rp}) at the measured upper frequency and the model
parameters are collected in $h=\{M,\alpha,Q\}$. We write $\ln\bar{\mathcal{L}}$ for the maximum
attained. The parameter space is sampled with a Metropolis--Hastings MCMC simulation of $\mathcal{O}(10^{5})$ iterations, adopting uniform broad priors
\begin{equation}
M\in[0,10]\,\Msun,\qquad \alpha\in[-15,15],\qquad Q\in[-20,20]\ {\rm km}.
\label{eq:priors}
\end{equation}
The prior on $\alpha$ is wide enough to include the values inferred from galactic rotation curves and
cluster dynamics, and the prior on $Q$ spans the range in which the $r^{-2}$ term is comparable to
the $r^{-1}$ term at the ISCO for a stellar-mass object. Sampling is unconstrained in sign for both
$\alpha$ and $Q$; only $Q^{2}$ enters Eq.~(\ref{eq:lapse}), so the posterior in $Q$ is symmetric
about zero by construction and the sign reported in the tables reflects the branch in which the chain
settled. For the microquasar sample a fourth parameter, the emission radius $r$, is sampled as well,
since the black hole data consist of a single frequency pair per source and the radius cannot be
eliminated as it is in the neutron star case.

MCMC sampling is used rather than a grid search because the parameter space is neither
low-dimensional nor separable. A probabilistic walk concentrates its effort in high-likelihood
regions while retaining a finite probability of exploring lower-likelihood ones, which is what allows
disconnected modes to be found, and its cost scales far more gently with dimension than any uniform
tessellation. Both properties matter here: Section \ref{isec7} shows that the posterior is genuinely
multimodal for most of the neutron star sources, and a grid coarse enough to be affordable in three
dimensions would have resolved at most one of the modes.

Models are compared with the deviance information criterion (DIC) \citep{Kunz:2006,Liddle:2007},
\begin{equation}
{\rm DIC}=2\left\langle-2\ln \mathcal{L}\right\rangle+2\ln\bar{\mathcal{L}},
\label{eq:dic}
\end{equation}
in which $\langle\ \rangle$ denotes an average over the posterior. The model attaining the smallest
value, ${\rm DIC}_{0}$, is taken as the reference, and the remaining models are ranked by the
difference $\Delta={\rm DIC}-{\rm DIC}_{0}$. Following the usual reading, a model with
$0\le\Delta\le3$ is weakly excluded, one with $3<\Delta\le6$ is mildly excluded, and one with
$\Delta>6$ is strongly excluded. The first term in Eq.~(\ref{eq:dic}) measures the average fit
quality and the second rewards a sharply peaked posterior, so the criterion penalises parameters that
the data do not constrain. This is exactly the situation created by the degeneracy of
Eq.~(\ref{eq:map}), and the reader should keep in mind when reading the tables that a large $\Delta$
for a model with more free parameters may signal an unconstrained direction rather than a poor fit.

Four spacetimes are compared for every source. The Schwarzschild case, labeled S, has $\alpha=Q=0$
and a single free parameter. The full STVG case allows both $\alpha$ and $Q$ to vary. The restricted
case $\alpha=0$ is the RN spacetime, and the restricted case $Q=0$ is the
Schwarzschild--MOG spacetime, in which the only correction is the one carried by the vector source
charge. Each of these must pass both a statistical and a physical selection criterion, the latter
requiring that the recovered parameters respect Eq.~(\ref{eq:ec}) and that the fitted orbits lie
outside the ISCO.

\section{Numerical Analyses, Results and Discussion}\label{isec7}

We now present the outcome of the sampling for the twelve sources and interpret it. The neutron star
sample and the microquasar sample behave very differently, and the difference is informative rather
than accidental: the neutron star data extend to high upper frequencies where the four theoretical
curves separate, whereas the black hole data consist of isolated pairs at frequencies where the
curves still coincide. We therefore treat the two samples in turn, and close the section with the
parameter degeneracy that ties the results together.

\begin{table}[ht!]
\centering
\footnotesize
\setlength{\tabcolsep}{10pt}
\renewcommand{\arraystretch}{1.42}
\begin{tabular}{|l|l|r|r|r|r|r|r|}
\hline
\textbf{X-ray binary} & \textbf{Model} & \multicolumn{3}{c|}{\textbf{MCMC best-fit model parameters}} & \multicolumn{3}{c|}{\textbf{Statistical criteria}}\\
\hline
& & $\boldsymbol{M/\Msun}$ & $\boldsymbol{\alpha}$ & $\boldsymbol{Q}$ \textbf{(km)} & $\boldsymbol{-\ln\bar{\mathcal{L}}}$ & \textbf{DIC} & $\boldsymbol{\Delta}$\\
\hline\hline
Cir X-1 & S & $2.224^{+0.029}_{-0.029}$ & -- & -- & $125.84$ & $254$ & $146$\\
\hline
 & STVG & $3.206^{+1.319}_{-0.720}$ & $0.95^{+0.71}_{-0.49}$ & $7.41^{+1.27}_{-1.90}$ & $51.89$ & $109$ & $1$\\
\hline
 & & $0.331^{+0.095}_{-0.063}$ & $7.49^{+2.80}_{-1.50}$ & $0$ & $116.47$ & $239$ & $131$\\
\hline
 & & $6.664^{+0.244}_{-0.233}$ & $0$ & $12.78^{+0.47}_{-0.46}$ & $51.89$ & $108$ & $0$\\
\hline
GX 5--1 & S & $2.161^{+0.010}_{-0.010}$ & -- & -- & $200.33$ & $403$ & $3$\\
\hline
 & STVG & $2.973^{+0.525}_{-0.480}$ & $-0.39^{+0.21}_{-0.11}$ & $-0.01^{+1.92}_{-1.86}$ & $197.84$ & $401$ & $1$\\
\hline
 & & $2.794^{+0.346}_{-0.249}$ & $-0.36^{+0.15}_{-0.06}$ & $0$ & $197.84$ & $400$ & $0$\\
\hline
 & & $2.171^{+0.033}_{-0.027}$ & $0$ & $0.07^{+0.78}_{-0.80}$ & $200.34$ & $405$ & $5$\\
\hline
GX 17+2 & S & $2.077^{+0.001}_{-0.001}$ & -- & -- & $1819.02$ & $3888$ & $3759$\\
\hline
 & STVG & $2.254^{+0.266}_{-0.216}$ & $1.13^{+0.30}_{-0.17}$ & $-4.26^{+0.36}_{-0.35}$ & $62.44$ & $135$ & $6$\\
\hline
 & & $<0.15$ & $>5.0$ & $0$ & $991.39$ & $2016$ & $1887$\\
\hline
 & & $4.945^{+0.037}_{-0.035}$ & $0$ & $-8.28^{+0.07}_{-0.07}$ & $62.44$ & $129$ & $0$\\
\hline
GX 340+0 & S & $2.102^{+0.003}_{-0.003}$ & -- & -- & $130.86$ & $264$ & $0$\\
\hline
 & STVG & $3.110^{+0.951}_{-0.703}$ & $-0.44^{+0.28}_{-0.12}$ & $-1.16^{+1.12}_{-2.52}$ & $130.17$ & $265$ & $1$\\
\hline
 & & $3.215^{+0.514}_{-0.540}$ & $-0.48^{+0.19}_{-0.10}$ & $0$ & $130.17$ & $265$ & $1$\\
\hline
 & & $2.117^{+0.101}_{-0.021}$ & $0$ & $-0.85^{+0.71}_{-0.54}$ & $130.86$ & $267$ & $3$\\
\hline
Sco X-1 & S & $1.965^{+0.001}_{-0.001}$ & -- & -- & $3887.17$ & $8518$ & $8231$\\
\hline
 & STVG & $3.355^{+0.095}_{-0.109}$ & $0.25^{+0.04}_{-0.03}$ & $6.66^{+0.15}_{-0.15}$ & $136.62$ & $291$ & $5$\\
\hline
 & & $<0.84$ & $>2.0$ & $0$ & $3134.63$ & $6298$ & $6012$\\
\hline
 & & $4.196^{+0.008}_{-0.027}$ & $0$ & $6.92^{+0.02}_{-0.05}$ & $136.62$ & $286$ & $0$\\
\hline
4U 1608--52 & S & $1.960^{+0.004}_{-0.004}$ & -- & -- & $235.83$ & $474$ & $347$\\
\hline
 & STVG & $2.719^{+0.670}_{-0.998}$ & $0.48^{+0.91}_{-0.27}$ & $4.70^{+0.64}_{-1.63}$ & $60.95$ & $127$ & $0$\\
\hline
 & & $<0.75$ & $>3.7$ & $0$ & $181.56$ & $372$ & $245$\\
\hline
 & & $4.131^{+0.070}_{-0.081}$ & $0$ & $6.81^{+0.13}_{-0.16}$ & $60.95$ & $127$ & $0$\\
\hline
4U 1728--34 & S & $1.734^{+0.003}_{-0.003}$ & -- & -- & $212.61$ & $427$ & $347$\\
\hline
 & STVG & $3.050^{+0.370}_{-0.315}$ & $0.27^{+0.16}_{-0.13}$ & $5.24^{+0.44}_{-0.43}$ & $37.99$ & $82$ & $1$\\
\hline
 & & $<0.50$ & $>3.7$ & $0$ & $169.84$ & $349$ & $268$\\
\hline
 & & $3.917^{+0.065}_{-0.074}$ & $0$ & $6.54^{+0.11}_{-0.15}$ & $37.99$ & $81$ & $0$\\
\hline
4U 0614+091 & S & $1.904^{+0.001}_{-0.001}$ & -- & -- & $842.97$ & $1690$ & $1399$\\
\hline
 & STVG & $3.746^{+0.368}_{-0.306}$ & $0.18^{+0.12}_{-0.09}$ & $6.43^{+0.48}_{-0.45}$ & $142.70$ & $292$ & $1$\\
\hline
 & & $<0.53$ & $>3.9$ & $0$ & $673.92$ & $1359$ & $1068$\\
\hline
 & & $4.496^{+0.045}_{-0.050}$ & $0$ & $7.55^{+0.08}_{-0.09}$ & $142.70$ & $291$ & $0$\\
\hline
\end{tabular}
\caption{MCMC results for the selected X-ray binaries hosting neutron stars. Columns list the
source, the model, the best-fit parameters $M$, $\alpha$ and $Q$ with $1\sigma$ errors, the
log-likelihood maximum as $-\ln\bar{\mathcal{L}}$, the DIC of Eq.~(\ref{eq:dic}) and its difference
$\Delta$ from the reference model. For each source the first STVG row has both parameters free, the
second has $Q=0$ and the third has $\alpha=0$.}
\label{tab:ns}
\end{table}

Table \ref{tab:ns} collects the neutron star results. The pattern is consistent across the sample.
For seven of the eight sources the charged models reach a log-likelihood maximum far above the
Schwarzschild one, and the DIC differences are correspondingly large, often in the hundreds and in
two cases in the thousands. Only GX 340+0 has the Schwarzschild model as its reference, and there the
competing models sit at $\Delta=1$ and $\Delta=3$, so nothing is excluded. The uncharged STVG
variant, in which the only modification is the vector source charge, behaves poorly for the atoll
sources and for Sco X-1: the chains drift to very small masses and very large couplings, hitting the
prior edge, with $\Delta$ in the hundreds or thousands. This is the expected consequence of the
degeneracy discussed below, since with $Q=0$ the two effective parameters are locked together as
$\Qeff^{2}=\alpha(1+\alpha)M^{2}$ and cannot be adjusted independently to match the data.

The masses inferred under the modified models are larger throughout the sample than those under
Schwarzschild. In the Schwarzschild fits the neutron star masses cluster between $1.7$ and
$2.2\,\Msun$, entirely consistent with the observed distribution and with the
Tolman--Oppenheimer--Volkoff limit \citep{Lamb:1998,Miller:1998,Trumper:2011,Cipolletta:2015,Breu:2016}.
Under STVG the same data return values up to $3.2\,\Msun$ for Cir X-1 in the fully free fit and up to
$6.7\,\Msun$ in the $\alpha=0$ branch. Taken at face value this would place several of these objects
above the maximum mass supported by any causal equation of state. The correct reading is more
guarded, and Eq.~(\ref{eq:map}) supplies it: what the timing data constrain is the effective mass
$\Meff=(1+\alpha)M$, and the inflation of $M$ is arithmetic rather than physical whenever $\alpha$ is
allowed to be large and negative or the branch with $\alpha=0$ is selected. In STVG the repulsive
vector interaction does genuinely oppose the gravitational attraction at short range, so
configurations above the general-relativistic limit are not excluded in principle; but the QPO data
by themselves do not establish them.

\begin{figure}[ht!]
\centering
\includegraphics[width=0.487\textwidth,clip]{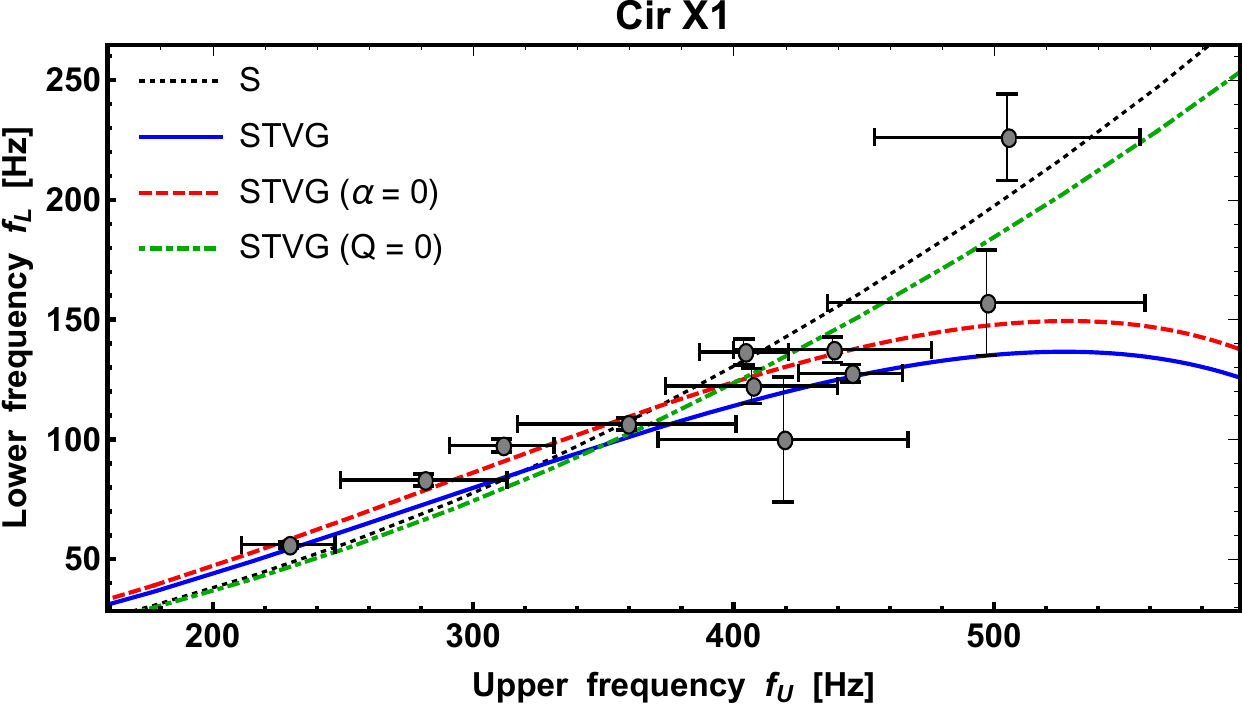}\hfill
\includegraphics[width=0.487\textwidth,clip]{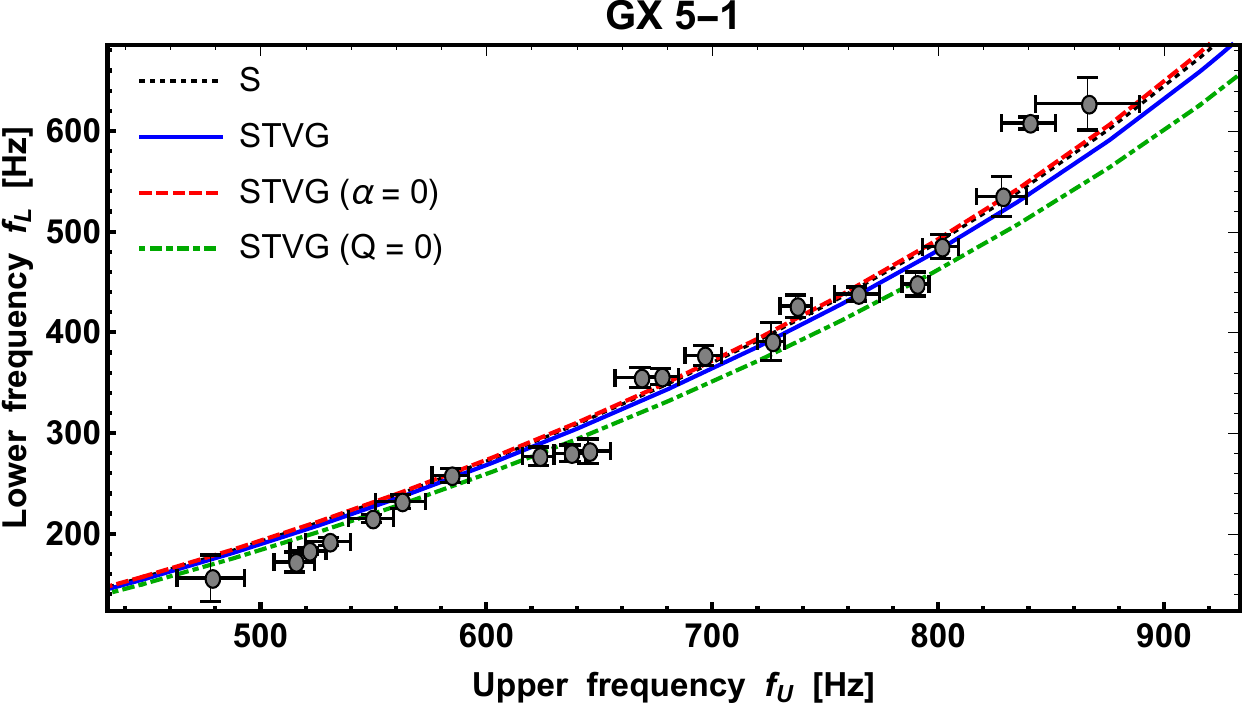}\\[3pt]
\includegraphics[width=0.487\textwidth,clip]{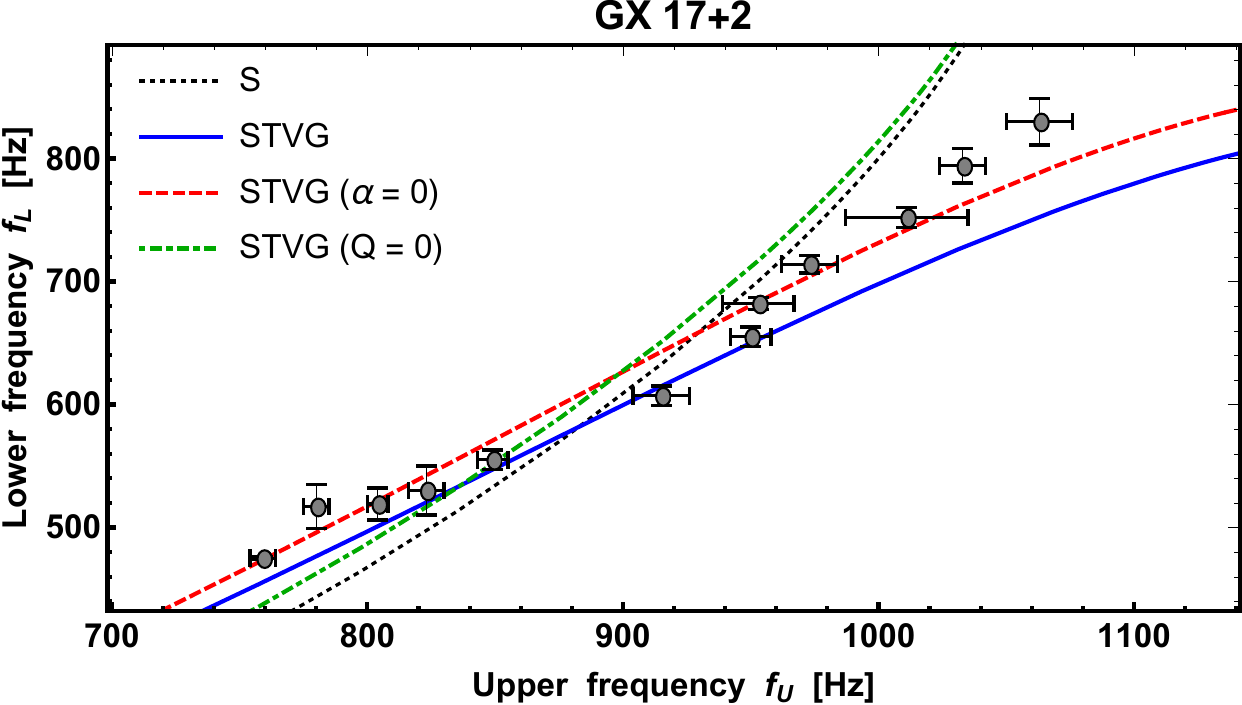}\hfill
\includegraphics[width=0.487\textwidth,clip]{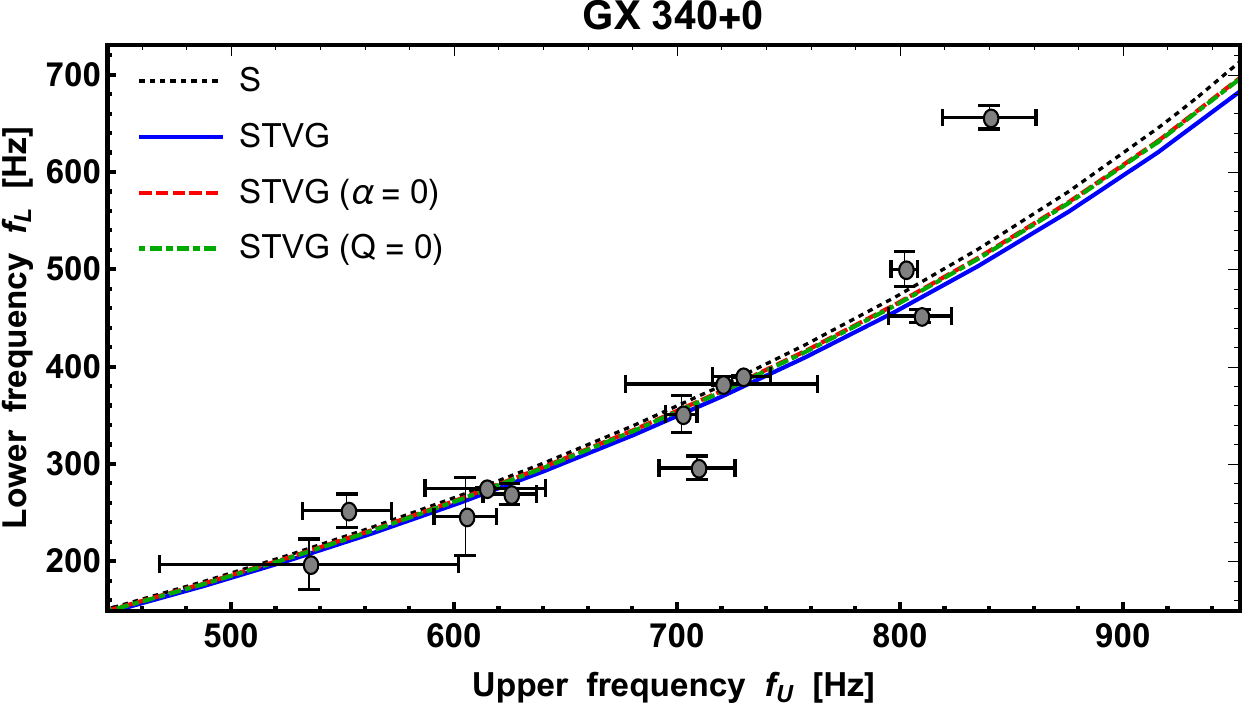}\\[3pt]
\includegraphics[width=0.487\textwidth,clip]{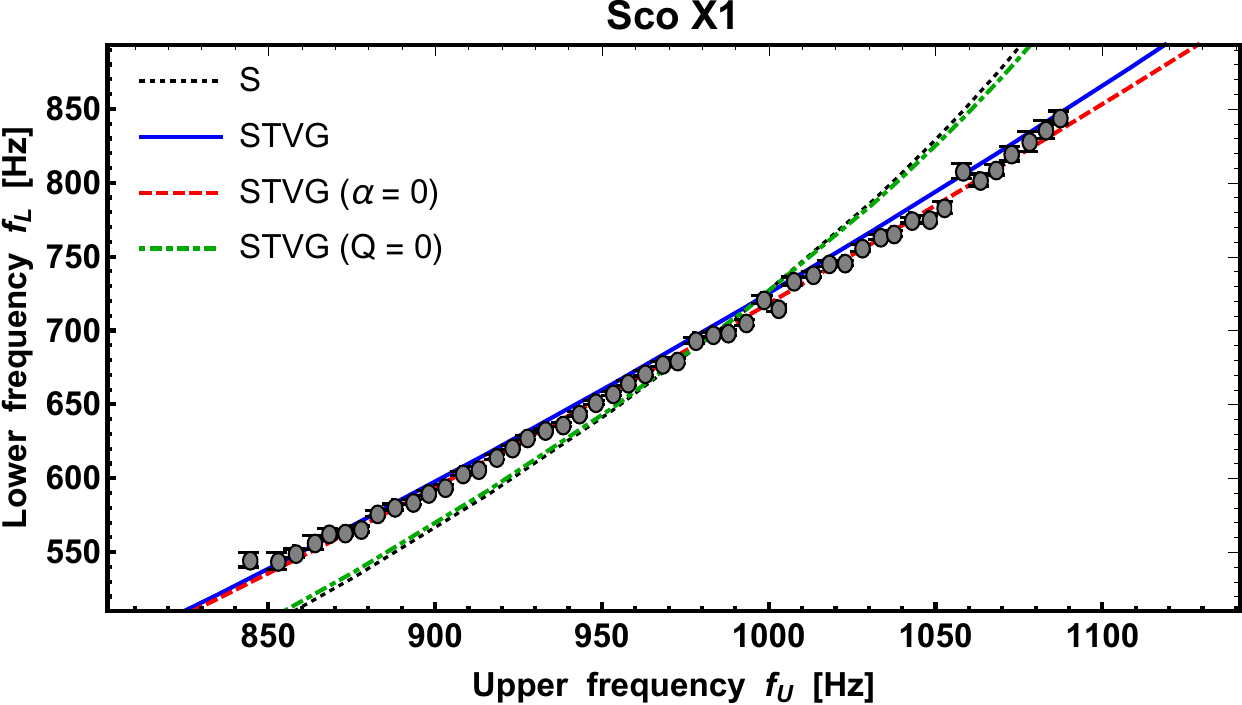}\hfill
\includegraphics[width=0.487\textwidth,clip]{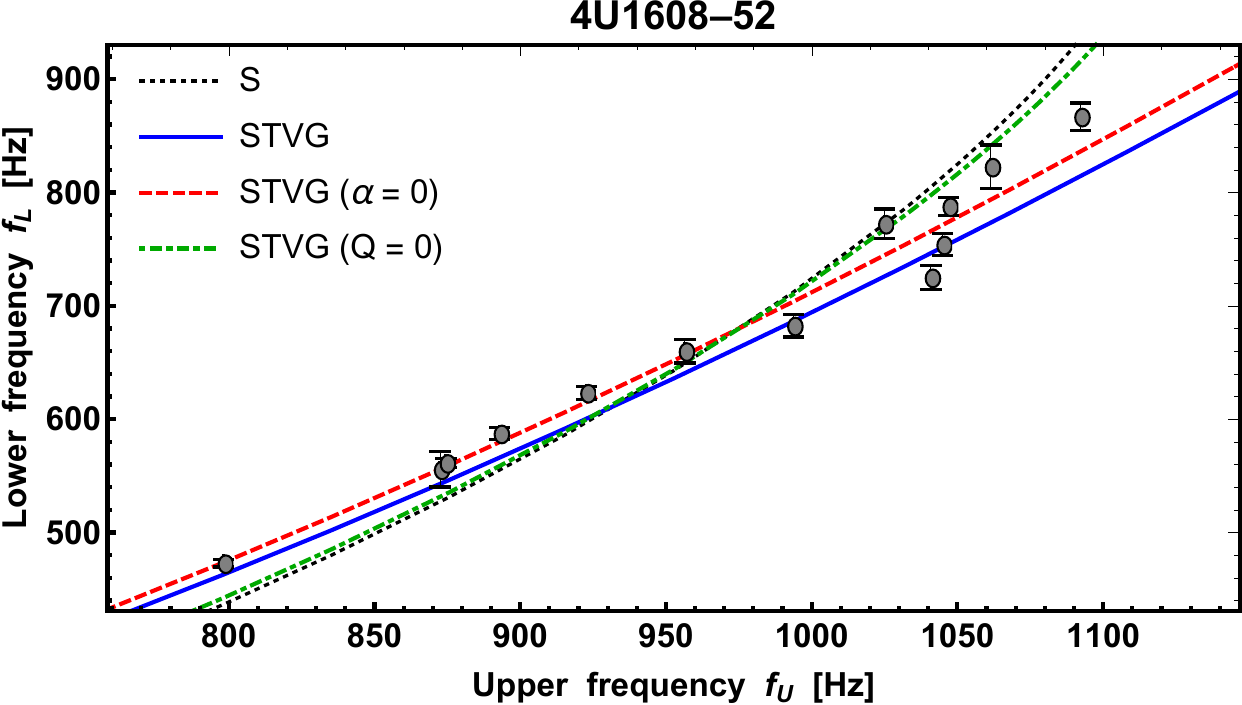}\\[3pt]
\includegraphics[width=0.487\textwidth,clip]{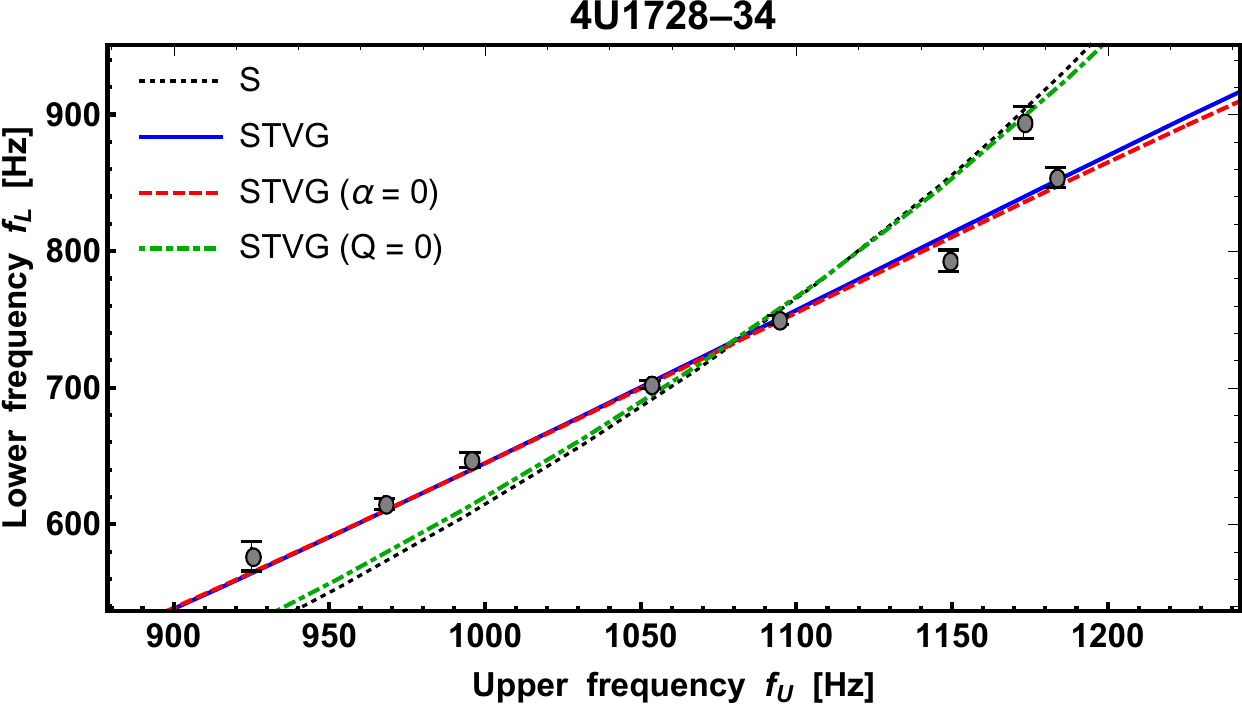}\hfill
\includegraphics[width=0.487\textwidth,clip]{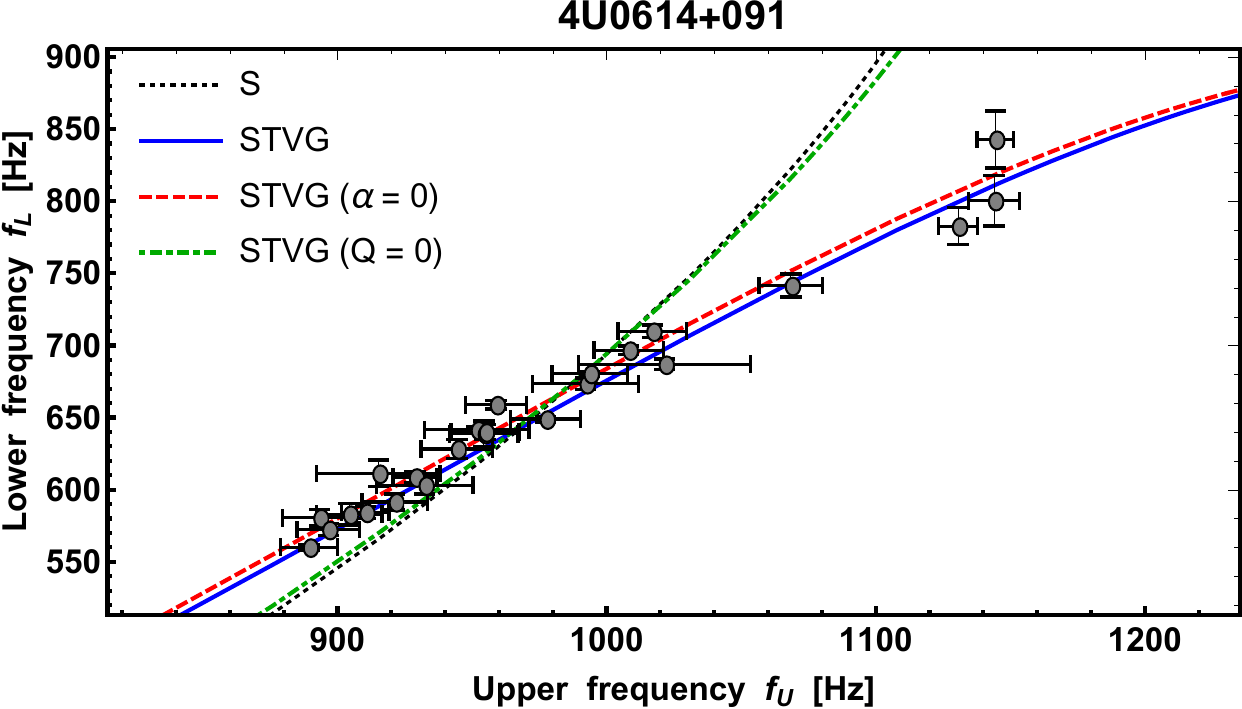}
\caption{The $(f_{\rm L},f_{\rm U})$ pairs of the binaries hosting neutron stars and the best-fitting curves for STVG (solid blue), STVG with $\alpha=0$ (dashed red), STVG with $Q=0$ (dot-dashed green), and Schwarzschild (dotted black).\\}
\label{fig:freqfits}
\end{figure}

Figure \ref{fig:freqfits} shows the data and the four best-fitting curves for each neutron star
source. The contour plots of the corresponding best-fit parameters are displayed in Appendix~\ref{app:A}. At low and moderate upper frequencies the four curves are nearly indistinguishable, which is
why the black hole sample discussed below carries so little discriminating power. As $f_{\rm U}$
increases they separate. The Schwarzschild curve rises steeply and overshoots $f_{\rm L}$ at the high
end, because in that geometry the turnover occurs only at the ISCO frequency of a
$\sim2\,\Msun$ object, well above the observed range. The full STVG curve levels off and bends
downward, matching the flattening seen in Cir X-1 and 4U 1728--34, and it gives the closest match to
the high-frequency data. The two restricted curves fall between these extremes: the $\alpha=0$ curve
removes the coupling enhancement but retains the repulsive term and behaves much like the full STVG
curve, whereas the $Q=0$ curve removes the repulsive term and tracks the Schwarzschild curve. Both
$\alpha$ and $Q$ thus act as damping parameters that prevent the predicted lower frequency from
growing without bound, and the extra field content is what allows the theoretical relation to follow
the data at the top of the observed range.

Counting sources, the STVG curve provides the best or joint-best match for six of the eight panels,
namely Cir X-1, Sco X-1, 4U 1608--52, 4U 1728--34 and 4U 0614+091, with GX 5--1 and GX 340+0 tied
among the models. The $Q=0$ curve is only marginally worse than the full STVG curve for the Z
sources but clearly worse for the atoll sources. GX 17+2 is the one panel in which the Schwarzschild
curve is not badly beaten in shape, although its DIC is far larger. Taken together the sample favors
the charged models, and the improvement is concentrated where it should be, at $f_{\rm U}$ above
about $1$ kHz, in the sources whose inner discs are most strongly perturbed.

\begin{table}[ht!]
\centering
\footnotesize
\setlength{\tabcolsep}{10pt}
\renewcommand{\arraystretch}{1.42}
\begin{tabular}{|l|l|r|r|r|r|r|r|r|}
\hline
\textbf{X-ray binary} & \textbf{Model} & \multicolumn{4}{c|}{\textbf{Best-fit parameters}} & \multicolumn{3}{c|}{\textbf{Statistical criteria}}\\
\hline
& & $\boldsymbol{M/\Msun}$ & $\boldsymbol{\alpha}$ & $\boldsymbol{Q}$ \textbf{(km)} & $\boldsymbol{r}$ \textbf{(km)} & $\boldsymbol{-\ln\bar{\mathcal{L}}}$ & \textbf{DIC} & $\boldsymbol{\Delta}$\\
\hline\hline
XTE J1550--564 & S & $9.14^{+0.33}_{-0.37}$ & -- & -- & $63.68^{+1.29}_{-1.18}$ & $29.72$ & $64$ & $0$\\
\hline
 & STVG & $9.64^{+0.82}_{-0.97}$ & $-0.06^{+0.15}_{-0.12}$ & $-0.18^{+2.94}_{-2.74}$ & $63.47^{+1.60}_{-1.64}$ & $29.42$ & $68$ & $4$\\
\hline
 & & $9.57^{+0.81}_{-0.91}$ & $-0.07^{+0.14}_{-0.12}$ & $0$ & $63.47^{+1.60}_{-1.64}$ & $29.41$ & $65$ & $1$\\
\hline
 & & $9.22^{+0.30}_{-0.45}$ & $0$ & $0.24^{+2.74}_{-3.08}$ & $63.77^{+1.32}_{-1.28}$ & $29.72$ & $66$ & $2$\\
\hline
GRO J1655--40 & S & $6.03^{+0.11}_{-0.12}$ & -- & -- & $41.05^{+0.39}_{-0.37}$ & $50.74$ & $106$ & $0$\\
\hline
 & STVG & $5.72^{+0.42}_{-0.42}$ & $0.08^{+0.13}_{-0.10}$ & $0.17^{+1.22}_{-1.48}$ & $41.35^{+0.49}_{-0.53}$ & $50.18$ & $109$ & $3$\\
\hline
 & & $5.71^{+0.39}_{-0.41}$ & $0.08^{+0.12}_{-0.09}$ & $0$ & $41.32^{+0.47}_{-0.55}$ & $50.18$ & $107$ & $1$\\
\hline
 & & $6.04^{+0.11}_{-0.13}$ & $0$ & $-0.10^{+1.27}_{-1.09}$ & $41.06^{+0.42}_{-0.38}$ & $50.74$ & $108$ & $2$\\
\hline
GRS 1915+105 & S & $15.10^{+0.93}_{-1.02}$ & -- & -- & $105.09^{+3.43}_{-3.57}$ & $17.66$ & $39$ & $0$\\
\hline
 & STVG & $12.87^{+2.65}_{-1.90}$ & $0.26^{+0.35}_{-0.30}$ & $0.18^{+8.32}_{-7.44}$ & $107.94^{+4.29}_{-4.81}$ & $17.18$ & $42$ & $3$\\
\hline
 & & $12.43^{+2.83}_{-2.10}$ & $0.22^{+0.48}_{-0.23}$ & $0$ & $107.33^{+4.34}_{-4.58}$ & $17.09$ & $40$ & $1$\\
\hline
 & & $15.27^{+0.89}_{-1.19}$ & $0$ & $0.22^{+5.93}_{-6.78}$ & $105.45^{+3.40}_{-3.96}$ & $17.66$ & $41$ & $2$\\
\hline
H1743--322 & S & $11.36^{+0.56}_{-0.66}$ & -- & -- & $76.28^{+1.88}_{-2.10}$ & $13.08$ & $30$ & $0$\\
\hline
 & STVG & $11.51^{+3.04}_{-2.68}$ & $0.02^{+0.44}_{-0.32}$ & $-1.00^{+9.96}_{-8.09}$ & $77.00^{+2.78}_{-3.26}$ & $13.05$ & $35$ & $5$\\
\hline
 & & $11.46^{+2.19}_{-2.05}$ & $-0.06^{+0.35}_{-0.19}$ & $0$ & $76.18^{+2.56}_{-2.66}$ & $13.05$ & $32$ & $2$\\
\hline
 & & $11.53^{+0.82}_{-0.83}$ & $0$ & $0.36^{+6.12}_{-6.58}$ & $76.34^{+2.26}_{-2.32}$ & $13.08$ & $32$ & $2$\\
\hline
\end{tabular}
\caption{MCMC results for the selected X-ray binaries hosting black holes. Columns list the source,
the model, the best-fit values of $M$, $\alpha$, $Q$ and the emission radius $r$ with $1\sigma$
errors, the log-likelihood maximum as $-\ln\bar{\mathcal{L}}$, the DIC of Eq.~(\ref{eq:dic}) and its
difference $\Delta$. For each source the first STVG row has both parameters free, the second has
$Q=0$ and the third has $\alpha=0$.}
\label{tab:bh}
\end{table}

Table \ref{tab:bh} gives the microquasar results (the corresponding contour plots are displayed in Appendix~\ref{app:A}), and the message is the reverse of the neutron star
one. In all four systems the Schwarzschild model attains the minimum DIC, and every STVG variant
carries a penalty $\Delta\ge1$. The posterior intervals for $\alpha$ and $Q$ include zero at
$1\sigma$ in every case without exception, so the black hole data provide no statistically meaningful
evidence for a departure from GR. The log-likelihood maxima do improve slightly under
the modified models, but never by enough to offset the complexity penalty, and the inferred black
hole masses and characteristic emission radii are stable against the choice of model. The reason is
visible in Fig.~\ref{fig:rp}: at the frequencies at which these sources oscillate the four
theoretical relations have not yet separated, so the added parameters are unconstrained and the DIC
correctly reports them as such.

The four systems differ in how tightly they are constrained. GRO J1655--40 has the smallest emission
radius, $r\simeq41$ km, the least massive black hole, $M\simeq5.71$ to $6.04\,\Msun$, and the
tightest bounds on every parameter, with $M=6.03^{+0.11}_{-0.12}\,\Msun$ in the Schwarzschild fit.
GRS 1915+105 is the most massive, $M=12.43$ to $15.27\,\Msun$, and has the largest spatial extent,
$r=105$ to $108$ km. XTE J1550--564 and H1743--322 are intermediate, with $M\simeq9.14$ to
$9.64\,\Msun$ and $r\simeq63.5$ km for the former and $M\simeq11.36$ to $11.53\,\Msun$ and
$r\simeq76$ to $77$ km for the latter. The uncertainties on the modified-gravity parameters are
largest for GRS 1915+105 and H1743--322, reaching $Q=-1.00^{+9.96}_{-8.09}$ km in the latter, which
is to say that the charge is entirely unconstrained.

The DIC penalty itself varies with how the parameters are restricted, and it does so regularly.
Model S is the most restricted baseline, carrying no free STVG parameter at all. The STVG variants
add structural freedom by letting both parameters float, by fixing $Q=0$, or by fixing $\alpha=0$.
Fully free fits attract the largest penalties, $\Delta=4$ for XTE J1550--564 and $\Delta=5$ for
H1743--322, while the singly restricted fits attract smaller ones, $\Delta=1$ to $2$. Since the
underlying fit quality is essentially identical across the four models, the ordering of $\Delta$ is
tracking the width of the posterior rather than the residuals, which is the intended behavior of
Eq.~(\ref{eq:dic}) and a direct measure of how little the data say about $\alpha$ and $Q$ in this
regime.

\begin{figure}[t!]
\centering
\includegraphics[width=\textwidth]{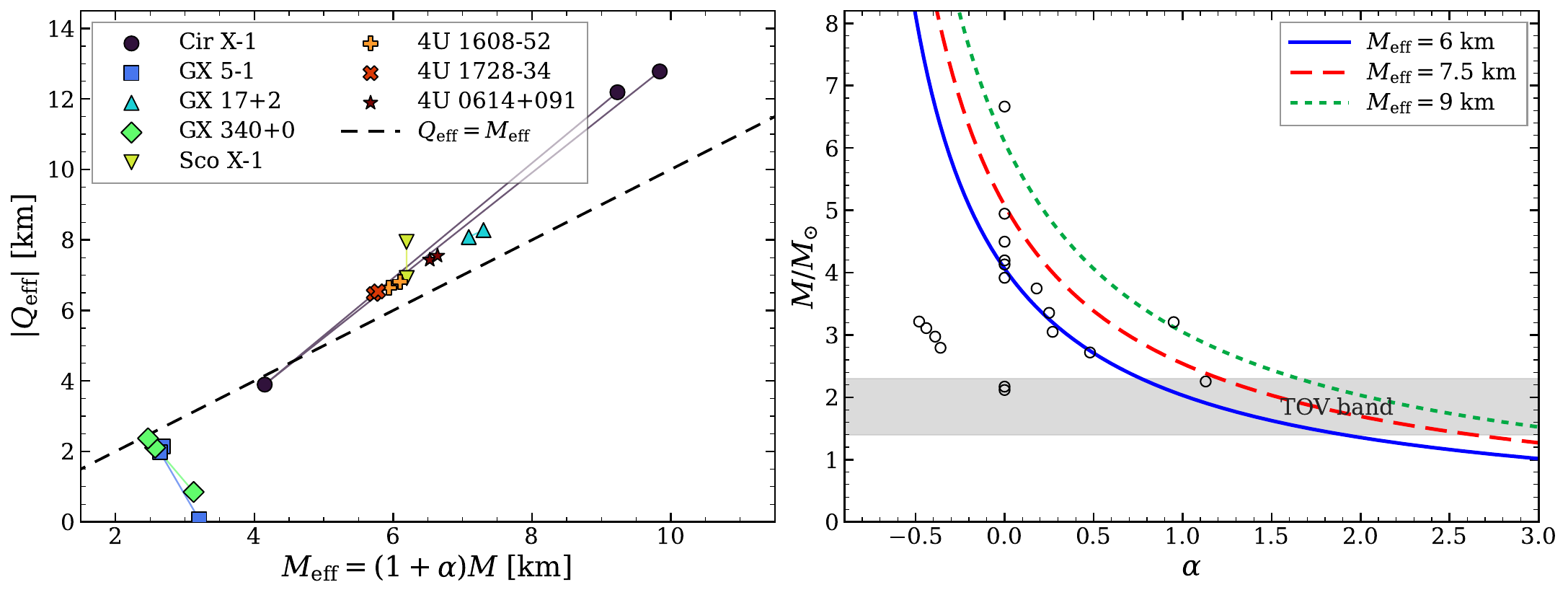}
\vspace{-0.3cm}\\
\caption{Left: the effective parameters of Eq.~(\ref{eq:map}) evaluated on every branch reported in
Table \ref{tab:ns}. Each symbol is one row of the table and branches belonging to the same source are
joined by a line; the dashed black line marks the extremal locus $\Qeff=\Meff$. Branches that share a
DIC collapse onto nearly the same point, whereas the branches with large $\Delta$ lie far away.
Right: contours of constant $\Meff$ in the $(\alpha,M)$ plane for $\Meff=6$, $7.5$ and $9$ km, with
the open circles marking the neutron star best fits of Table \ref{tab:ns} and the shaded band showing
the range of maximum masses allowed by causal equations of state. Moving along a contour changes the
inferred mass by a large factor while leaving every orbital frequency unchanged.}
\label{fig:degeneracy}
\end{figure}

The degeneracy anticipated in Section \ref{isec2} now becomes quantitative, and
Fig.~\ref{fig:degeneracy} makes it visible. Evaluating Eq.~(\ref{eq:map}) on the branches of
Table \ref{tab:ns} gives, for GX 17+2, $\left(\Meff,\Qeff\right)=(7.09,8.08)$ km on the fully free
branch against $(7.30,8.28)$ km on the $\alpha=0$ branch; for 4U 1728--34, $(5.72,6.47)$ km against
$(5.78,6.54)$ km; for 4U 0614+091, $(6.53,7.44)$ km against $(6.64,7.55)$ km; and for 4U 1608--52,
$(5.94,6.64)$ km against $(6.10,6.81)$ km. In each of these cases the two branches carry the same
DIC to within one unit, and they agree in the effective plane to within a few percent, the residual
difference being consistent with the rounding of the quoted parameters. The multiple parameter modes
noted in the table are therefore not distinct physical solutions. They are separate points on a
single degenerate ridge, and the sampler has settled at different places on that ridge depending on
where the chain began. Where the two branches do disagree in the effective plane, as for the $Q=0$
rows of Cir X-1 and GX 5--1, the DIC difference is correspondingly large, which is exactly what one
expects if the effective parameters and not the bare ones determine the fit.

The right panel of Fig.~\ref{fig:degeneracy} draws the practical consequence. Along a contour of
constant $\Meff$ the predicted frequency pair is invariant, yet the bare mass ranges from below one
solar mass at large $\alpha$ to well above the Tolman--Oppenheimer--Volkoff band at $\alpha$ near
$-1$. Any statement about a neutron star mass extracted from QPO timing in this theory is therefore a
statement about $\Meff$ together with a prior on $\alpha$, and the apparently dramatic masses in
Table \ref{tab:ns} are a reflection of the flat prior we adopted rather than a measurement. Breaking
the degeneracy requires an observable that is not built from equatorial test-particle motion in the
same metric. Two candidates suggest themselves. A shadow measurement determines $b_{\rm c}$ through
Eq.~(\ref{eq:bcrit}), but since $b_{\rm c}$ is also a function of $\Meff$ and $\Qeff$ alone it does
not help by itself; what it does provide is an independent handle on the same two combinations, and
combining it with the timing data would tighten them jointly. A dynamical mass from the binary orbit does not help
either, since a Keplerian probe at any radius measures $\Meff$ rather than $M$. What does separate
the two is the finite range of the vector field, which the solution adopted here sets to zero:
restoring it makes the enhancement scale dependent, so that a binary orbit and an inner accretion
disc sample different effective couplings. For GRO J1655--40 and GRS 1915+105, whose companion
masses are known from optical spectroscopy, quantifying that difference is the natural first
step.

Placing these results in context, the pure Schwarzschild metric serves as a useful reference but
consistently fails to account for the curvature of the $f_{\rm L}(f_{\rm U})$ relation at
$f_{\rm U}>1$ kHz, where the inner disc dynamics are most strongly affected by the magnetic field and
the gravitational field of the host neutron star, particularly in the highly variable sources Cir X-1
and GX 17+2. Rotating and deformed exteriors such as Kerr or Hartle--Thorne address the same
shortfall by adding frame dragging and a mass quadrupole
\citep{Boshkayev:2026,Boshkayev:2024,Urbancova:2019}, and regular black hole spacetimes address it by
modifying the near-horizon geometry \citep{Boshkayev:2023,Boshkayev:2023a}. STVG addresses it through
the enhanced gravitational coupling, which smooths the high-frequency end of the relation and yields
a markedly better description of the upper-kilohertz scatter in the neutron star low-mass X-ray
binaries. That three physically distinct mechanisms produce a comparable improvement is itself
informative, and it means that the present bounds on $\alpha$ and $Q$ should be read as bounds
conditional on the assumption of staticity. For the black hole candidates GRO J1655--40 and
GRS 1915+105 the mass estimates are stable across all four models, at $M\simeq5.7$ to
$6.0\,\Msun$ and $M\simeq12.4$ to $15.2\,\Msun$ respectively, in agreement with earlier analyses of
charged and rotating black holes fitted to the same sources \citep{Nishonov:2025,Oteev:2026}.

\section{Conclusion}\label{isec8}

We have analyzed the QPO data of eight neutron star sources and four black hole binaries within the STVG, obtaining bounds on the mass $M$, the coupling
$\alpha$ and the charge $Q$ by comparing the Schwarzschild solution with the full STVG solution and
with its two restrictions $\alpha=0$ and $Q=0$. The comparison rested on MCMC sampling of the RP likelihood and on the DIC, and it was preceded by a complete analytic treatment of the geometry in which every quantity entering the
model was checked against its RN and Schwarzschild limits.

The analytic part produced three results that stand independently of the fits. The charged STVG
geometry is isometric to a RN geometry with $\Meff=(1+\alpha)M$ and
$\Qeff^{2}=(1+\alpha)(\alpha M^{2}+Q^{2})$, so that every observable built from test-particle motion
depends on two combinations of the three parameters rather than on the three separately. The horizon
structure follows from this in closed form, with the extremal configuration occurring at $|Q|=M$
whatever the value of the coupling, and the effective source turns out to be traceless and
Maxwell-like, so that the weak and null energy conditions require only $\alpha>-1$ together with
$\alpha M^{2}+Q^{2}\ge0$. The radial epicyclic frequency is fixed by three independent
checks: the form given in Eq.~(\ref{eq:omr2compact}) reduces to both known limits and generates on
its vanishing the same ISCO cubic that the closed-form marginally stable radius satisfies. The fits
and figures reported here are reproduced by this expression.

Turning to the fits, the neutron star sample favors the charged solutions decisively. Seven of the eight sources return a DIC for the modified models that lies far below
the Schwarzschild value, the difference reaching several thousand for Sco X-1 and GX 17+2, and the
improvement is concentrated at upper frequencies above one kilohertz where the four theoretical
relations separate. Both $\alpha$ and $Q$ act to damp the predicted lower frequency and to produce
the flattening that the data display in Cir X-1 and 4U 1728--34, and the uncharged variant performs
poorly precisely because fixing $Q=0$ locks the two effective parameters together. Under Schwarzschild
the inferred masses stay within the customary range of $1.7$ to $2.2\,\Msun$, while the modified
parameterizations push the inferred value as high as $3.2\,\Msun$ for Cir X-1 in the fully free fit
and higher still on the $\alpha=0$ branch. We have argued that this inflation should not be read as
evidence for neutron stars above the Tolman--Oppenheimer--Volkoff limit. What the timing data measure
is the effective mass, and the bare mass is recovered from it only after a prior on $\alpha$ has been
imposed. The repulsive vector interaction of the theory does genuinely permit stable configurations
above the general-relativistic limit, but establishing them requires an independent measurement.

The microquasar sample tells the opposite story and is equally instructive. In all four systems the
Schwarzschild model attains the minimum criterion, every modified variant carries a penalty, and the
posterior bounds on $\alpha$ and $Q$ contain zero at one standard deviation without exception, so
that the present black hole data show no statistically significant evidence for a departure from GR. The inferred masses and characteristic emission radii are stable against the
choice of model, with GRO J1655--40 the tightest constrained at $M=5.71$ to $6.04\,\Msun$ and
$r\simeq41$ km, and GRS 1915+105 the largest and least certain at $M=12.43$ to $15.27\,\Msun$ and
$r=105$ to $108$ km. The modified models do lower the negative log-likelihood slightly, but never by
enough to offset the complexity cost, and the ordering of the penalties across the restricted fits
tracks the width of the posterior rather than the residuals. This is the expected outcome when the
data lie at frequencies where the competing relations have not yet diverged.

Two limitations frame these conclusions. The first is the degeneracy, which is a property of the
solution and not of the analysis, and which means that the quoted intervals on $\alpha$ and $Q$
separately are prior-dominated even where the intervals on their effective combinations are tight.
The second is staticity. A rotating central object splits the vertical epicyclic frequency from the
orbital one and introduces nodal precession, and since frame dragging and a mass quadrupole are known
to reproduce a comparable improvement in the same sources, part of what we have attributed to
$\alpha$ and $Q$ may be absorbing rotation. Taken together with the black hole results, the honest
summary is that STVG offers a viable route to supporting larger effective masses through repulsive
vector corrections, while the compact-binary QPO data currently in hand continue to support standard GR as the statistically most economical description.

Future tasks follow naturally in the near term. The most direct is to break the degeneracy by restoring the finite
range of the vector field, which makes the coupling scale dependent and so allows a binary orbit and
an inner disc to be compared; GRO J1655--40 and GRS 1915+105, whose companion masses are known from
optical spectroscopy, are the obvious first targets. A second task is to relax staticity. A rotating STVG black hole is
already known, but the slowly rotating charged exterior carrying a mass quadrupole, the STVG
counterpart of the Hartle--Thorne construction, is not. Building it to second order in the angular
velocity and refitting the eight neutron star sources would show how much of the preference for
$\alpha\neq0$ is instead absorbing frame dragging and quadrupole deformation. A third is to extend the analysis beyond the
relativistic precession model to the epicyclic resonance and diskoseismic families, since a
modification of the metric that improves one mechanism need not improve the others, and consistency
across mechanisms would strengthen any claim considerably. It would also be worth computing the
quasinormal spectrum of the corrected geometry and checking the eikonal correspondence against the
Lyapunov exponent reported in Fig.~\ref{fig:maps}, which would connect the timing bounds to
ringdown observations. Finally, the shadow radius of Eq.~(\ref{eq:bcrit}) and the deflection angle of
Eq.~(\ref{eq:defl}) provide a second, independent route to the same effective parameters, and a joint
analysis of horizon-scale imaging together with kilohertz timing for a single source would tighten
both far more than either can alone.

\section*{Acknowledgments}

\.I.S. is grateful to Eastern Mediterranean University, T\"UB\.ITAK, ANKOS and SCOAP3 for their
support, and acknowledges the networking support of COST Actions CA22113, CA21106, CA23130, CA21136
and CA23115.

\section*{Data Availability Statement}

The QPO frequency pairs analyzed in this work are available in the published
sources cited in Section \ref{isec1}. All derived quantities, together with the scripts that generate
the figures, are available from the corresponding author on reasonable request.

\bibliographystyle{unsrtnat_linkizzet}
\bibliography{finalref}

\appendix

\section{Contour plots}\label{app:A}

Hereby, we report the contour plots of the MCMC posteriors of the STVG model parameters obtained in Section~\ref{isec7}.
These are compared with the outcomes from the Schwarzschild spacetime.

Figure~\ref{fig:cont1} portrays the contour plots corresponding to the best-fit results of Table~\ref{tab:ns}, obtained for the eight X-ray binaries hosting neutron stars. For graphical display purposes, some contours are cut at convenient ranges, implying that some of the peaks of the 1D log-likelihoods appearing at the edges of the boxes are just interpolation artifacts. For the exact constraints, the reader is invited to refer to Table~\ref{tab:ns}.

Figure~\ref{fig:cont2} shows the contour plots corresponding to the best-fit results of Table~\ref{tab:bh}, obtained for the four microquasars considered in this work.
In these cases, all contours are well constrained within the displayed ranges, thus no artificial peaks of the 1D log-likelihoods appear at the edges of the boxes.

\begin{figure}[t!]
\centering

{\hfill
\includegraphics[width=0.45\textwidth,clip]{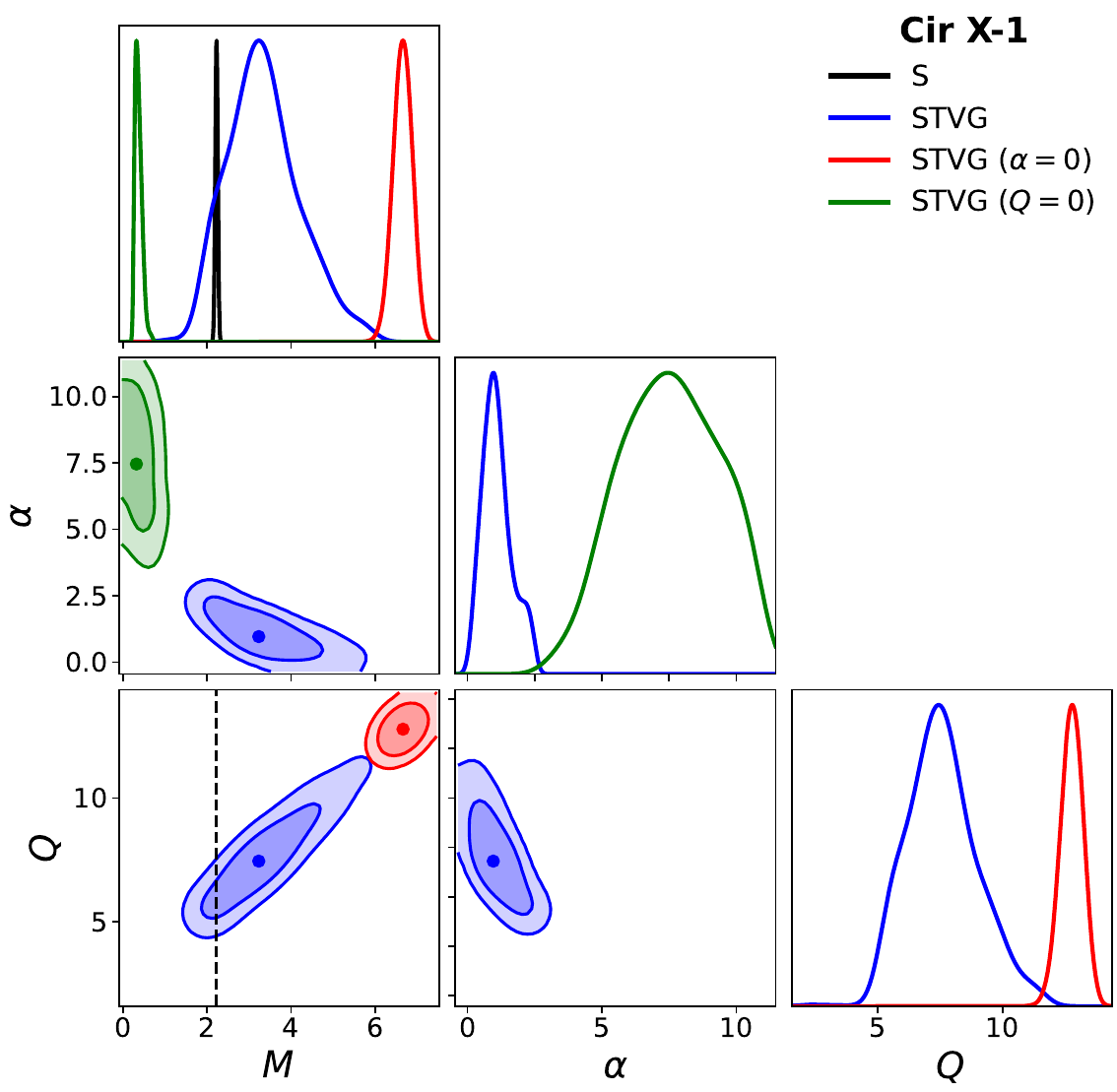}
\hfill
\includegraphics[width=0.45\textwidth,clip]{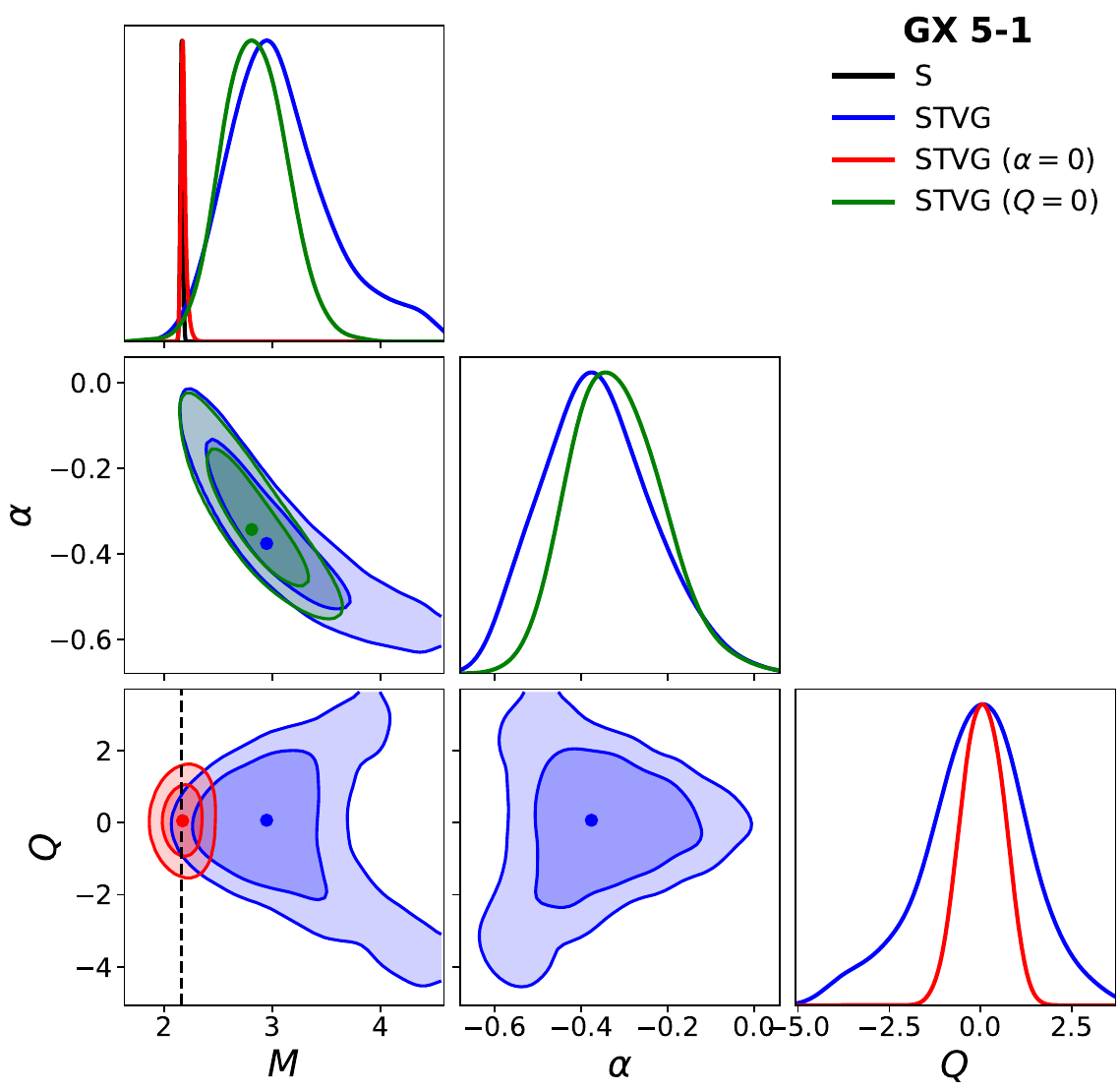}
\hfill}

{\hfill
\includegraphics[width=0.45\textwidth,clip]{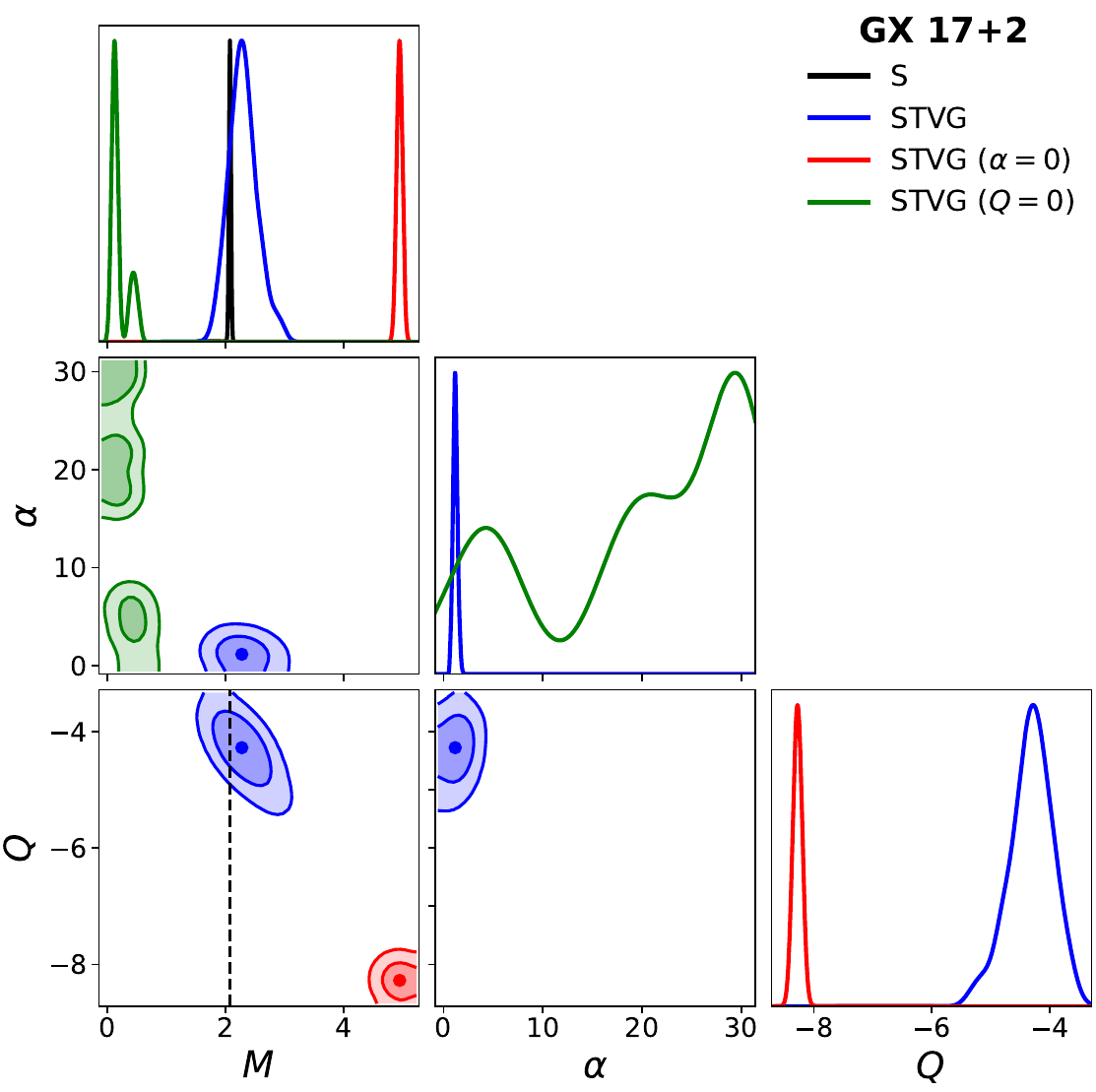}
\hfill
\includegraphics[width=0.45\textwidth,clip]{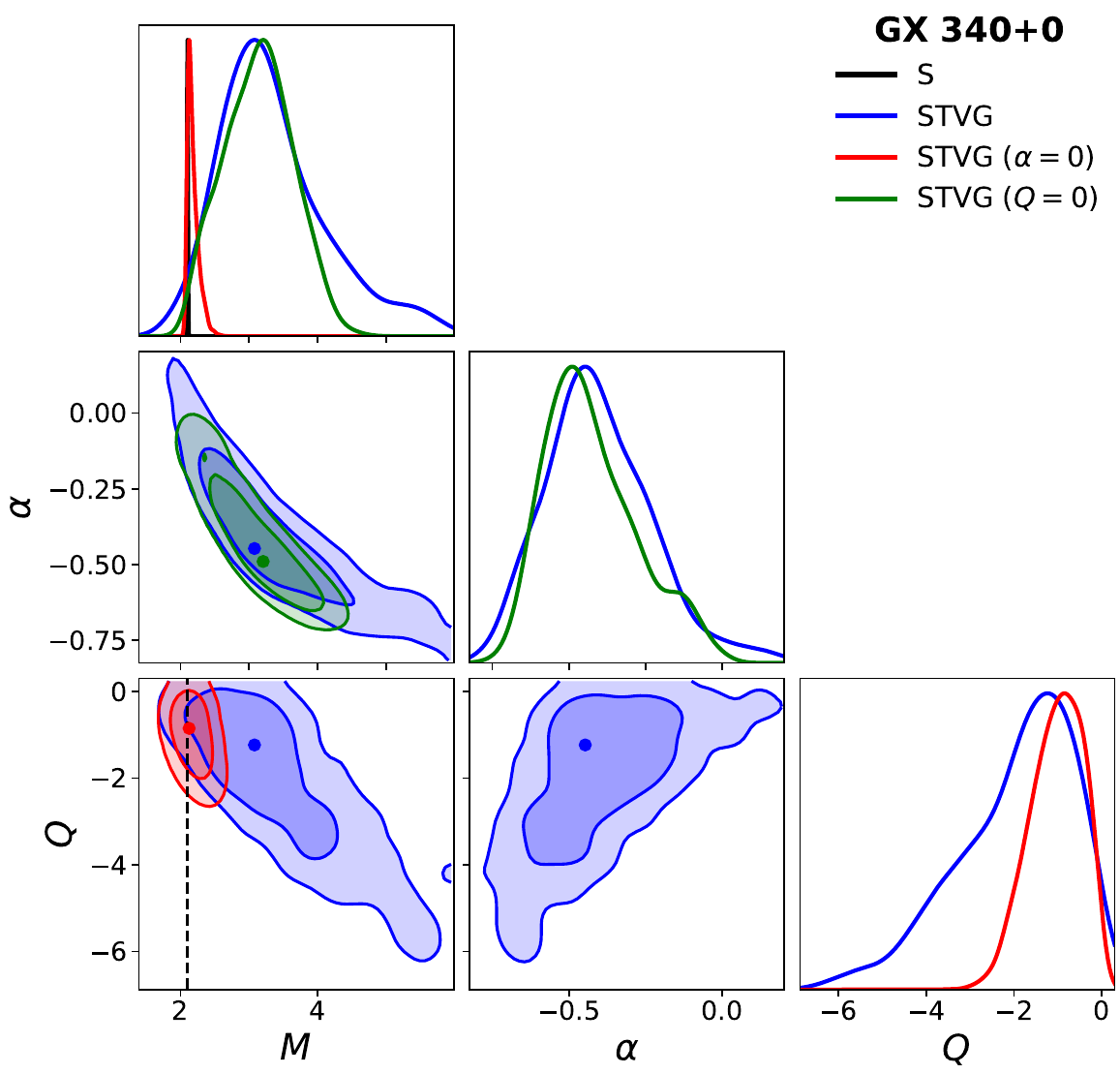}
\hfill}

\caption{MCMC posteriors of the Schwarzschild and the STVG models for the X-ray binaries hosting neutron stars. The color choices are the same as in Figure~\ref{fig:freqfits}; the dark (light) shaded areas correspond to $1\sigma$ ($2\sigma$) confidence levels. The dashed vertical lines mark the values of the mass parameter for the Schwarzschild model.}
\label{fig:cont1}
\end{figure}

\begin{figure}[t!]
\ContinuedFloat
\centering

{\hfill
\includegraphics[width=0.45\textwidth,clip]{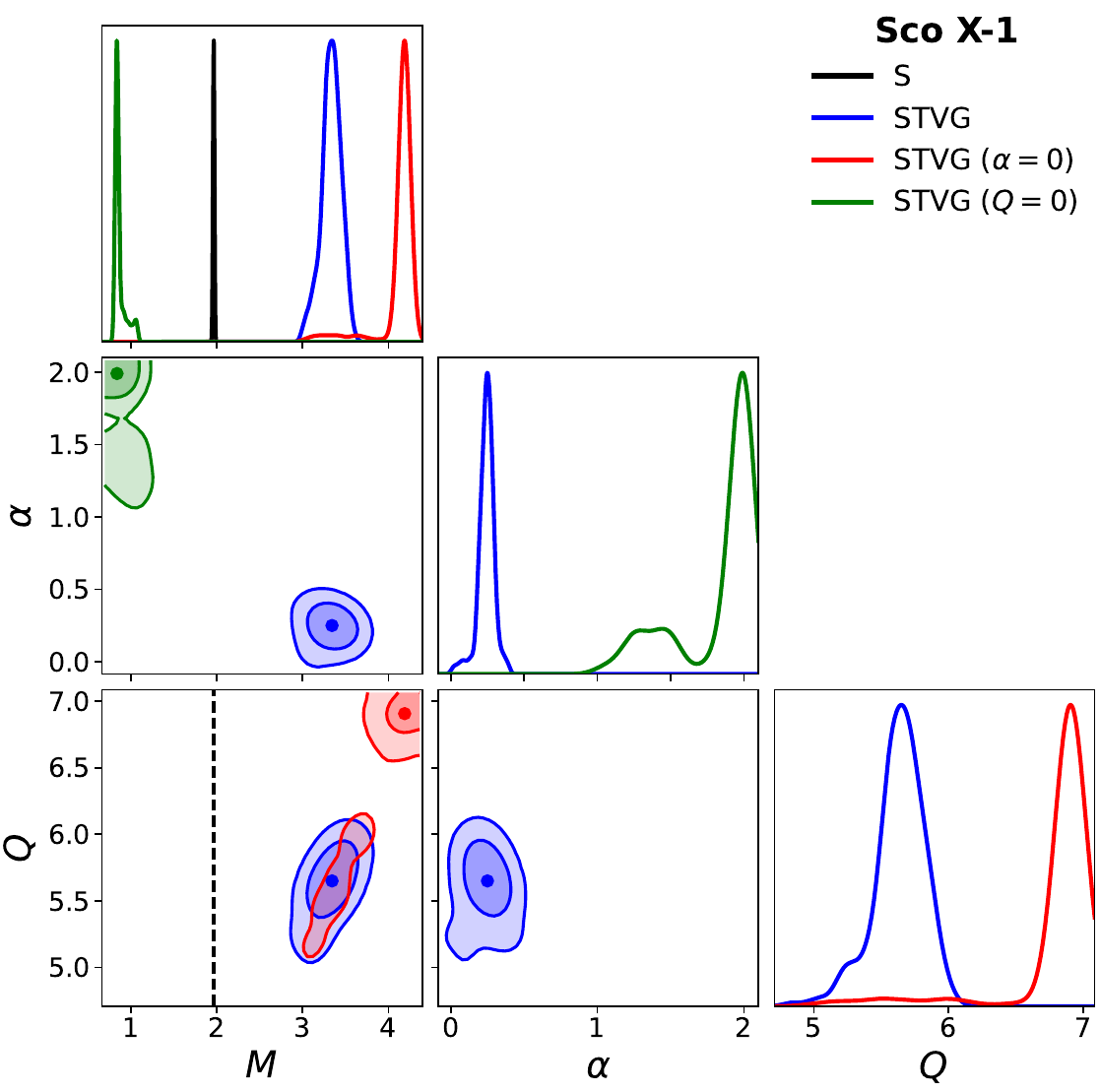}
\hfill
\includegraphics[width=0.45\textwidth,clip]{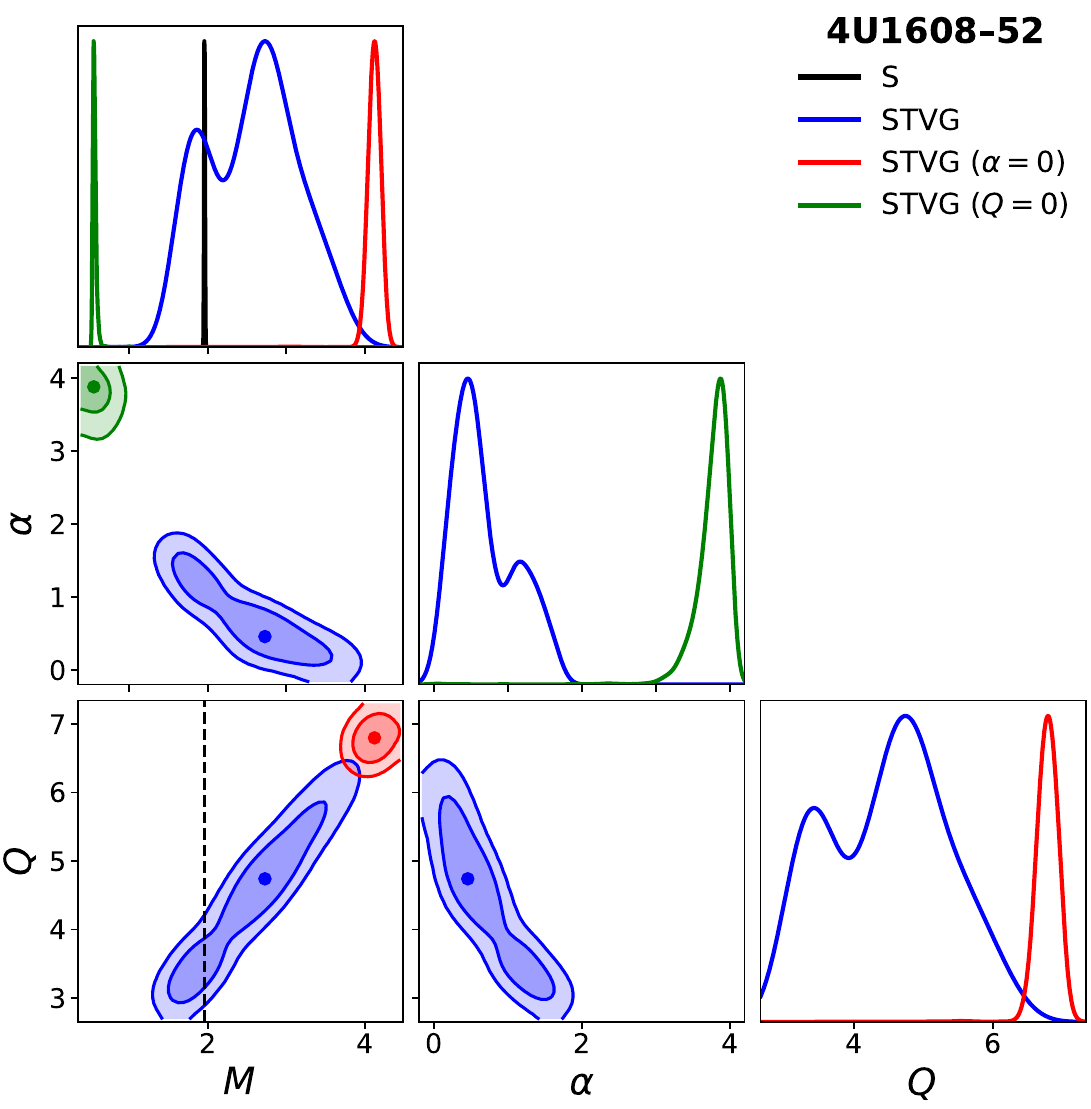}
\hfill}

{\hfill
\includegraphics[width=0.45\textwidth,clip]{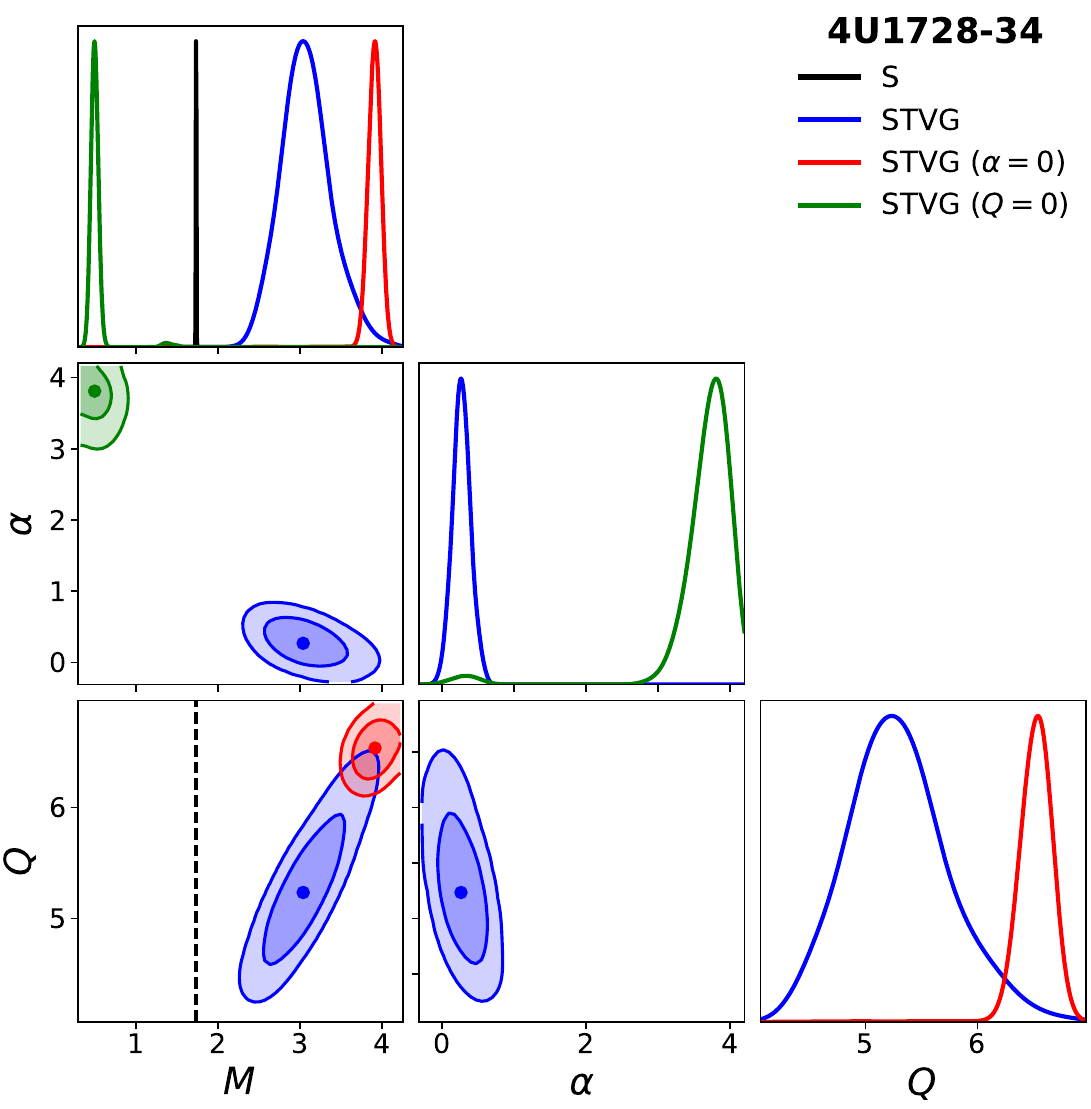}
\hfill
\includegraphics[width=0.45\textwidth,clip]{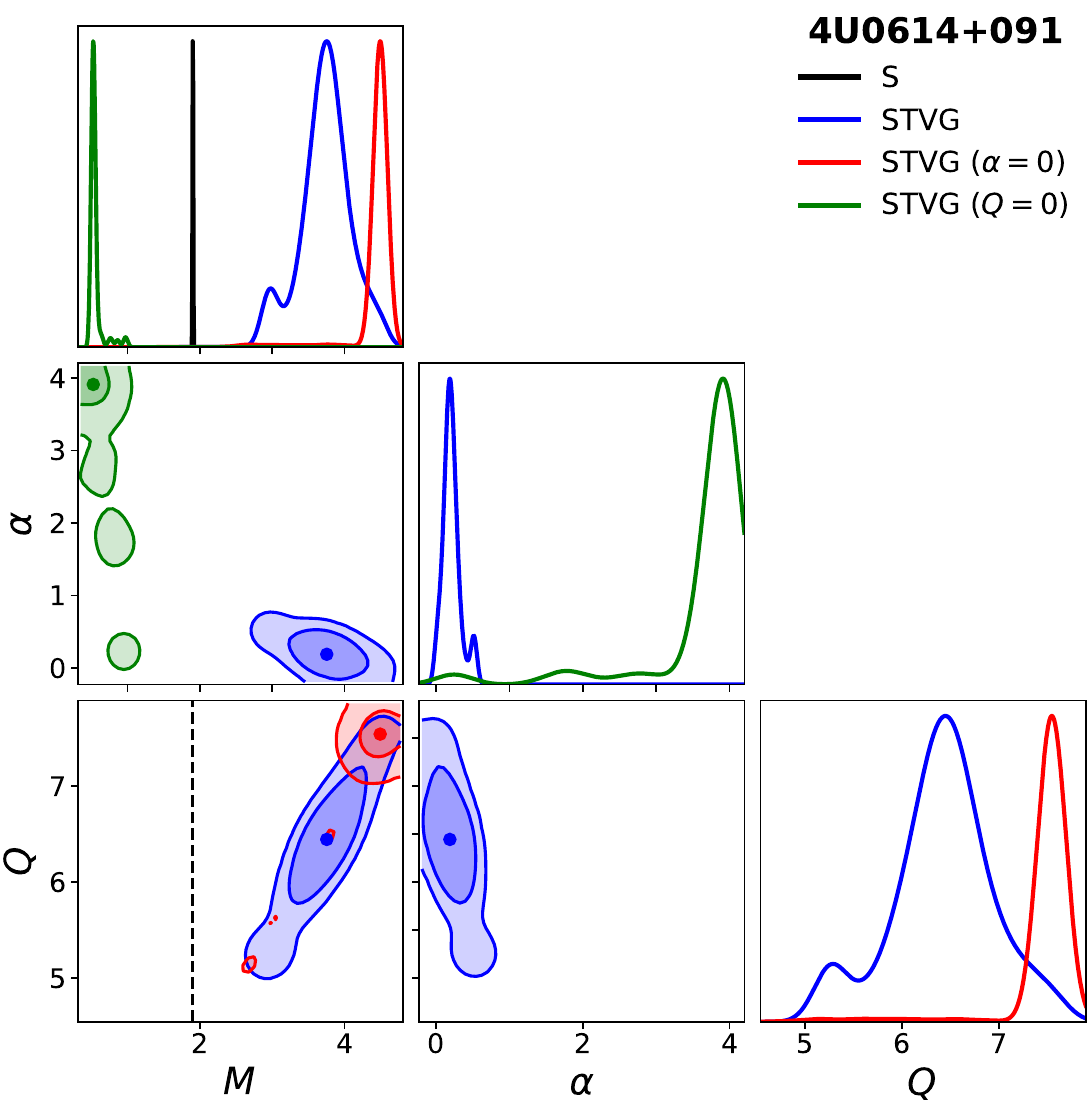}
\hfill}

\caption{Continued.}
\end{figure}

\begin{figure}[t!]
\centering

{\hfill
\includegraphics[width=0.49\textwidth,clip]{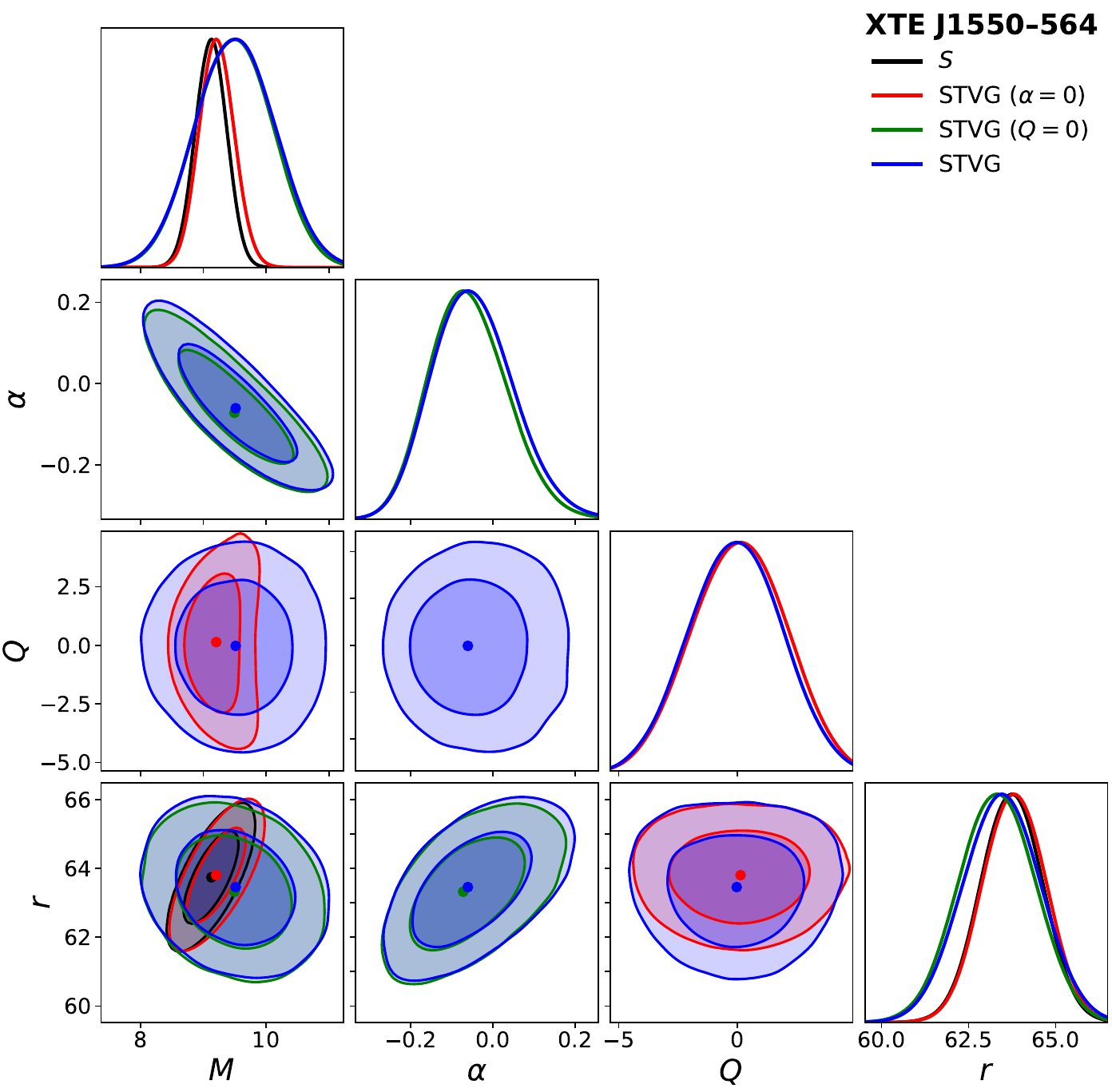}
\hfill
\includegraphics[width=0.49\textwidth,clip]{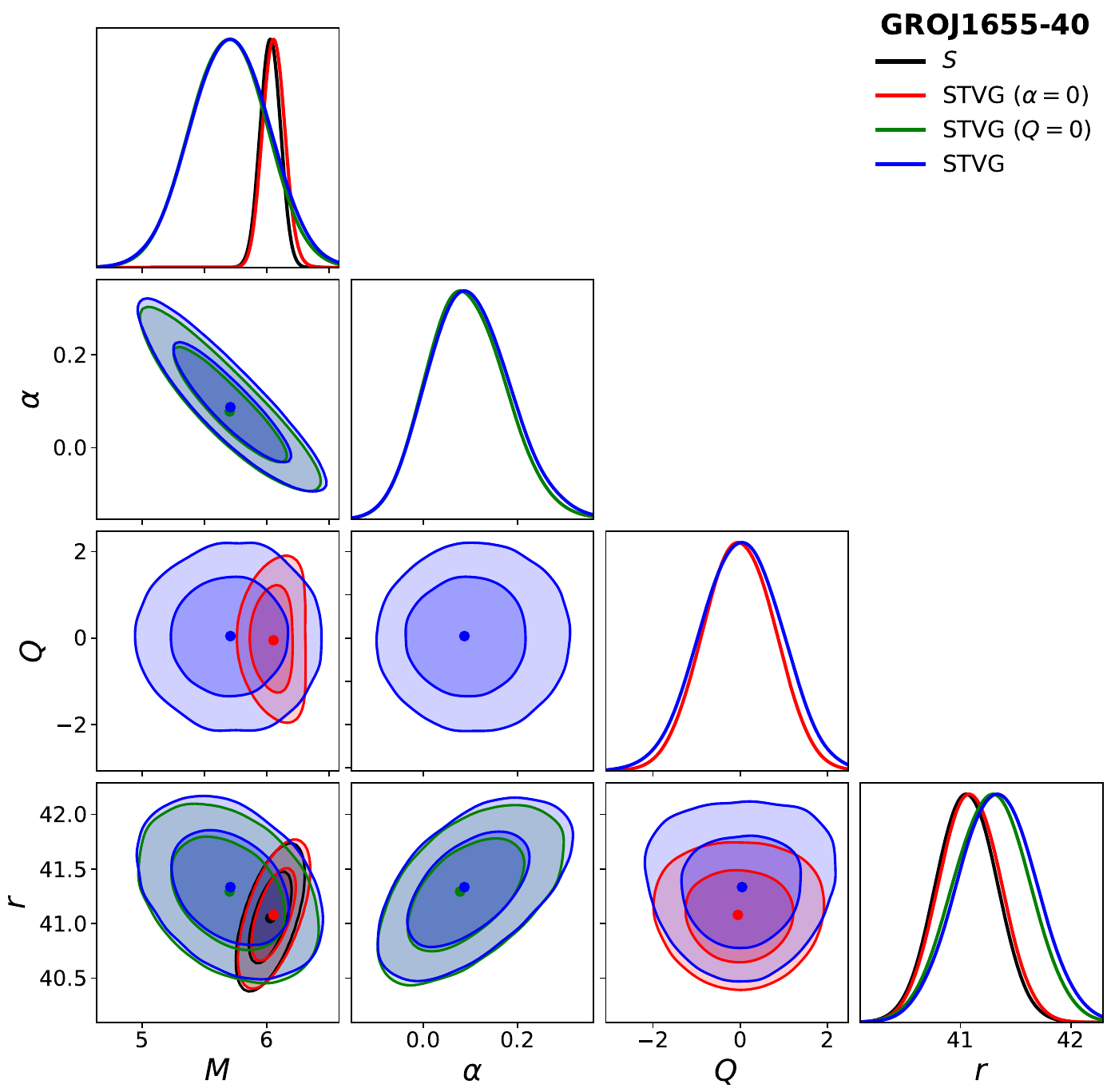}
\hfill}

{\hfill
\includegraphics[width=0.49\textwidth,clip]{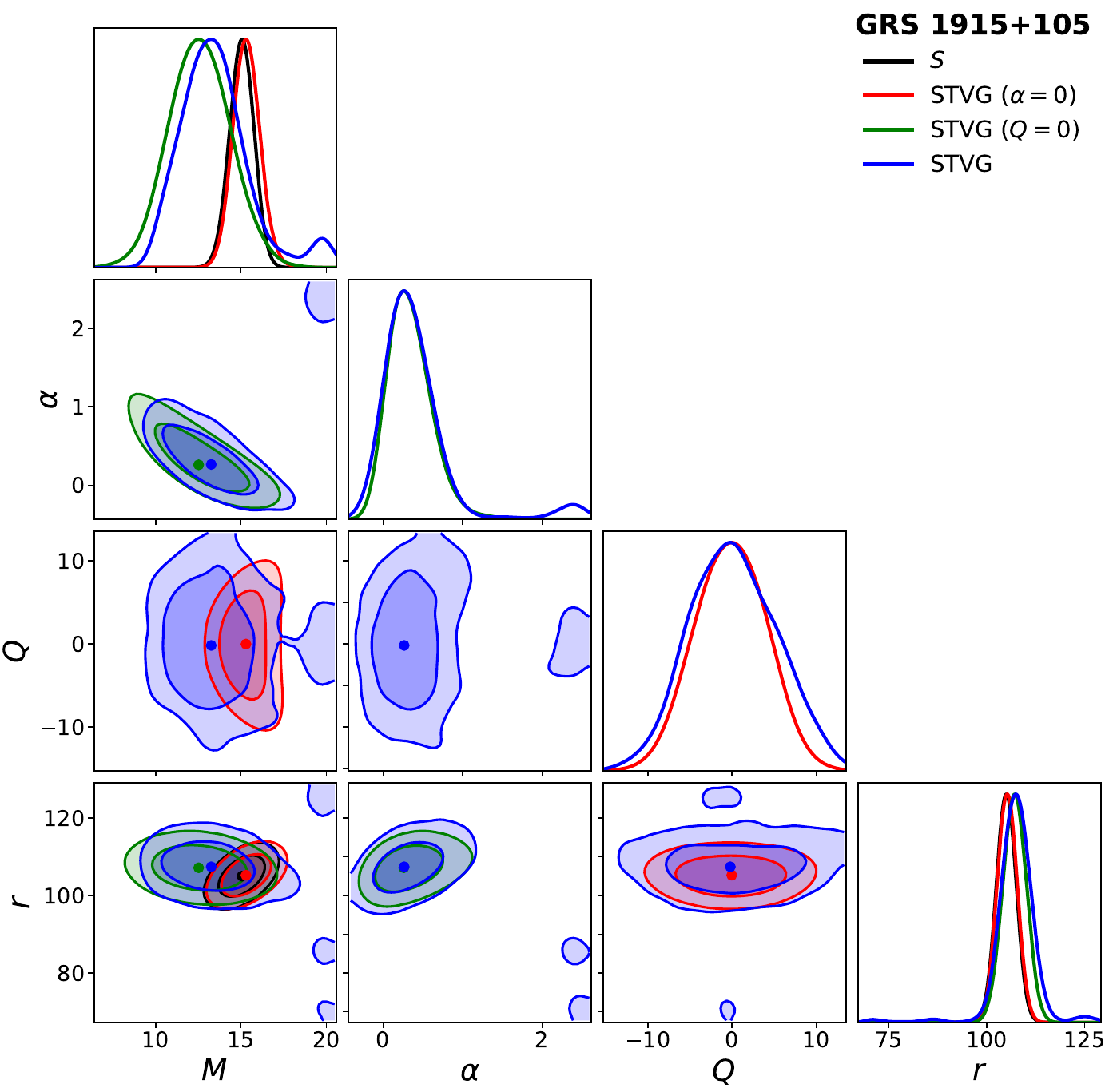}
\hfill
\includegraphics[width=0.49\textwidth,clip]{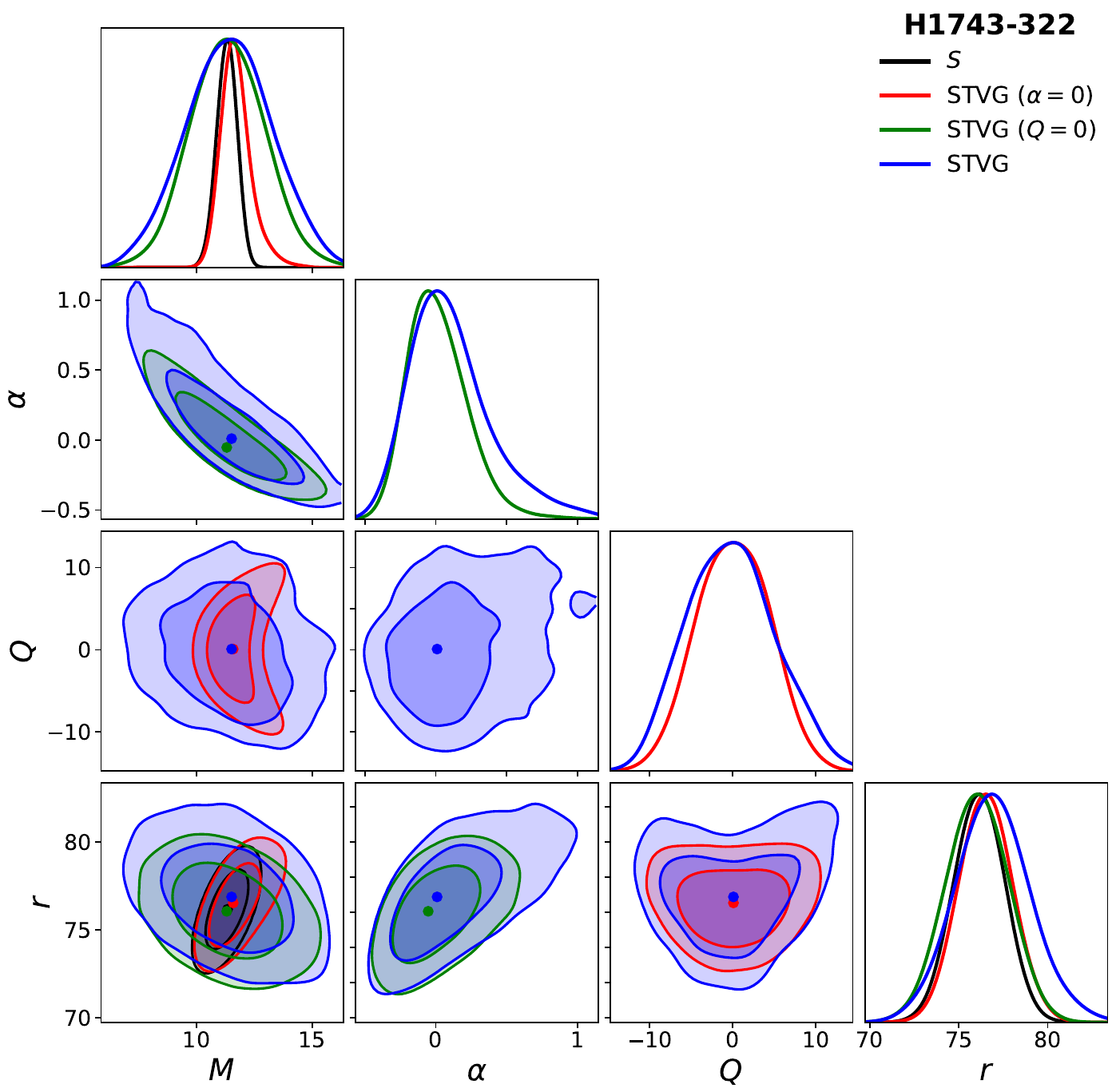}
\hfill}

\caption{MCMC posteriors of the Schwarzschild and the STVG models for microquasars. The color choices are the same as in Figure~\ref{fig:cont1}; the dark (light) shaded areas correspond to $1\sigma$ ($2\sigma$) confidence levels.}
\label{fig:cont2}
\end{figure}

\end{document}